\documentclass[jgrga]{agu2001}
\usepackage{graphicx}
\usepackage{amssymb}
\newcommand\Alfven{Alfv\'en }
\newcommand{\V}[1]{\mathbf{#1}} 
 
\newcommand{\figref}[1]{Figure~\ref{#1}}
\newcommand{\secref}[1]{\S~\ref{#1}}
\newcommand{\eqref}[1]{equation~(\ref{#1})}
\newcommand{\eqsref}[2]{equations~(\ref{#1})~and~(\ref{#2})}
\newcommand{\exref}[1]{(\ref{#1})}
\newcommand{\bea}{\begin{eqnarray}} 
\newcommand{\eea}{\end{eqnarray}} 
\newcommand{\beq}{\begin{equation}} 
\newcommand{\eeq}{\end{equation}} 
\newcommand{\lt}{\left}
\newcommand{\rt}{\right}
\newcommand{\kperp}{k_\perp}
\newcommand{\kpar}{k_\parallel}
\newcommand{\tomega}{\tilde\omega} 
\newcommand{\tgamma}{\tilde\gamma} 
	\newcommand{\domega}{\delta\tomega} 
\newcommand{\tOmega}{\tilde\Omega}

\usepackage{color}

\authorrunninghead{HOWES ET AL.}

\titlerunninghead{DISSIPATION OF THE SOLAR WIND}

\authoraddr{G. G. Howes,
Department of Astronomy, University of
California, 601 Campbell Hall,  Berkeley, CA 94720, USA.
(ghowes@astro.berkeley.edu)}

\begin{document}

\title{A Model of Turbulence in Magnetized Plasmas: 
Implications for the Dissipation Range in the Solar Wind}

\authors{G.~G. Howes, \altaffilmark{1}
 S.~C. Cowley, \altaffilmark{2, 5} W. Dorland, \altaffilmark{3}
 G.~W. Hammett, \altaffilmark{4} E. Quataert, \altaffilmark{1} and
 A.~A. Schekochihin \altaffilmark{5}}

\altaffiltext{1}{Department of Astronomy, University of
California, Berkeley, CA, USA.}
\altaffiltext{2}{Department of Physics and Astronomy, UCLA,
 Los Angeles, CA, USA.}
 \altaffiltext{3}{Department of Physics, IREAP, and Center for Scientific Computing and
Mathematical Modeling, University of Maryland, College Park, MD, USA.}
\altaffiltext{4}{Princeton Plasma Physics Laboratory, Princeton, NJ, USA.}
\altaffiltext{5}{Department of Physics, Imperial College, London, UK}

\begin{abstract}
This paper studies the turbulent cascade of magnetic energy in weakly
collisional magnetized plasmas. A cascade model is presented, based on
the assumptions of local nonlinear energy transfer in wavenumber
space, critical balance between linear propagation and nonlinear
interaction times, and the applicability of linear dissipation rates
for the nonlinearly turbulent plasma.  The model follows the nonlinear
cascade of energy from the driving scale in the MHD regime, through
the transition at the ion Larmor radius into the kinetic \Alfven wave
regime, in which the turbulence is dissipated by kinetic processes.
The turbulent fluctuations remain at frequencies below the ion
cyclotron frequency due to the strong anisotropy of the turbulent
fluctuations, $k_\parallel \ll k_\perp$ (implied by critical balance).
In this limit, the turbulence is optimally described by gyrokinetics;
it is shown that the gyrokinetic approximation is well satisfied for
typical slow solar wind parameters. Wave phase velocity measurements
are consistent with a kinetic \Alfven wave cascade and {\em not} the
onset of ion cyclotron damping. The conditions under which the
gyrokinetic cascade reaches the ion cyclotron frequency are
established. Cascade model solutions imply that collisionless damping
provides a natural explanation for the observed range of spectral
indices in the dissipation range of the solar wind. The dissipation
range spectrum is predicted to be an exponential fall off; the
power-law behavior apparent in observations may be an artifact of
limited instrumental sensitivity.  The cascade model is motivated by a
programme of gyrokinetic simulations of turbulence and particle
heating in the solar wind.
\end{abstract}

\begin{article}

\section{Introduction}

In the study of turbulence in magnetized plasmas, the solar wind
provides a unique opportunity to characterize the nature of turbulent
plasmas by making detailed dynamical measurements through
\emph{in situ} satellite observations. But using the solar wind as 
a turbulence laboratory is complicated by the fact that the solar wind
is only weakly collisional, so a  kinetic description is necessary 
\citep{Marsch:1991,Marsch:2006}. A useful first step in connecting theoretical 
ideas to observational data over the range from the large
energy-containing scales of the turbulence to the small scales where
the turbulence is dissipated is to have a simple model combining
our understanding of magnetized turbulence at both fluid and kinetic
scales. This paper presents such a model. 

This paper follows \citet{Howes:2006} and \citet{Schekochihin:2007} as
part of a program to study kinetic turbulence in astrophysical plasmas
using nonlinear gyrokinetic simulations. Here we attempt to model the
observed magnetic energy spectrum in the solar wind using a minimal
number of ingredients---namely, finite Larmor radius effects and
kinetic damping via the Landau resonance. A lengthy evaluation of the
applicability of the gyrokinetic approximation to the solar wind
plasma is presented along with a discussion of the relevant
observational data. The gyrokinetic cascade model yields sensible
results when compared to the observational data, but much additional
physics exists and may prove to be important; nonlinear simulations
will play a critical role in unraveling the complex interactions
occurring in the turbulent solar wind.

\subsection{Turbulence in Magnetized Plasmas}
\label{sec:turb}
The ubiquity of turbulence in space and astrophysical plasmas---e.g.,
the solar corona and solar wind, accretion disks around black holes,
the interstellar medium---has driven a vast effort to understand and
characterize turbulence in a magnetized plasma. Progress in the
understanding of magnetohydrodynamic (MHD) turbulence has accelerated
steadily over the decades. The pioneering work of
\citet{Iroshnikov:1963} and \citet{Kraichnan:1965} identified two key
properties of MHD turbulence: first, even in a magnetized plasma with
no mean field, the magnetic field of large-scale fluctuations can
behave effectively as a mean field for smaller-scale fluctuations; and
second, only oppositely directed
\Alfven wave packets interact nonlinearly. They assumed
weak nonlinear interactions and an isotropic
(with respect to the direction of the mean magnetic field) cascade of
energy and derived an energy spectrum that
scaled as $k^{-3/2}$. Measurements of magnetized turbulence in
laboratory plasmas \citep{Robinson:1971,Zweben:1979,Montgomery:1981} 
and in the solar wind \citep{Belcher:1971} as well as the
results of numerical simulations \citep{Shebalin:1983,Oughton:1994} brought attention to the anisotropy
inherent in MHD turbulence in the presence of a mean field and  
inspired the early anisotropic theories of MHD turbulence
\citep{Montgomery:1981,Montgomery:1982,Shebalin:1983,Higdon:1984a}. Based on these ideas,
\citet{Sridhar:1994} constructed a perturbative theory of weak MHD 
turbulence in which energy cascades only to higher perpendicular
wavenumbers; but their neglect of the $k_\parallel = 0$
modes in resonant three-wave couplings provoked criticisms
\citep{Montgomery:1995,Ng:1996}, and in response refined theories of 
weak MHD turbulence emerged
\citep{Goldreich:1997,Ng:1997,Galtier:2000,Lithwick:2003}.

The anisotropic nature of weak turbulence implies that it
generates fluctuations that become increasingly strong as the cascade  
proceeds to higher perpendicular wavenumber; combining this concept with the ideas
of \citet{Higdon:1984a}, \citet{Goldreich:1995} proposed that
a state of strong turbulence is eventually reached that 
maintains a {\em critical balance} 
between the (parallel) linear propagation 
and (perpendicular) nonlinear interaction timescales.
They proceeded to develop a scaling theory of strong turbulence in incompressible
MHD, hereafter referred to as the GS theory. Later studies have
examined the effects of compressibility
\citep{Lithwick:2001,Cho:2003,Vestuto:2003}, 
imbalanced Els\"asser fluxes
\citep{Maron:2001,Cho:2002,Lithwick:2007}, and a host of other refinements
\citep{Oughton:2004,Oughton:2005,Zhou:2005,Boldyrev:2005,Dastgeer:2006}.

The GS theory for strong, incompressible MHD turbulence rests on two
central assumptions: the locality of interactions in the wavenumber
space \citep{Kolmogorov:1941}, and the GS conjecture of critical
balance.  These imply a one-dimensional kinetic energy spectrum
$E_k(k_\perp) \propto k_\perp^{-5/3}$ (see \secref{sec:model}).  Using
the consequent scaling for the nonlinear frequency $\omega_{nl} \simeq
\kperp v_\perp\propto\kperp^{2/3}$ and equating it with the linear
frequency for shear \Alfven waves, $\omega = \pm k_\parallel v_A$
(critical balance), one finds $k_\parallel \propto k_\perp^{2/3}$.
Therefore, the fluctuations become more anisotropic at high
wavenumbers and the turbulence deep in the inertial range consists of
nearly perpendicular modes with $k_\perp\gg k_\parallel$. Numerical
simulations of MHD turbulence with a dynamically strong mean field
appear to support these conclusions \citep{Cho:2000,Maron:2001}.

In real astrophysical plasmas, although the large scales at which the
turbulence is driven may be adequately described by MHD, the turbulent
fluctuations at the small-scale end of the inertial range often have
parallel wavelengths smaller than the ion mean free path, $k_\parallel
\lambda_{{\rm mfp}i} \gg 1$; therefore, a kinetic description of the plasma is 
required to capture the turbulent dynamics (a systematic way to do
this is given by \citet{Schekochihin:2007}).  It is, however, possible
to show rigorously that Alfv\'enic fluctuations are essentially fluid
in nature, satisfy the Reduced MHD (RMHD) equations
\citep{Strauss:1976}, and remain undamped until their cascade reaches
the perpendicular scale of the ion Larmor radius, $k_\perp \rho_i \sim
1$
\citep{Quataert:1998,Lithwick:2001,Schekochihin:2007}. 
In contrast, the compressible MHD fluctuations such as the slow, fast,
and entropy modes require a kinetic description even at large scales
and are damped both collisionally by parallel ion viscosity
\citep{Braginskii:1965} at scales $k_\parallel \lambda_{{\rm mfp}i} \sim 1$ 
and collisionlessly by the ion Landau damping 
\citep{Barnes:1966} at scales $k_\parallel \lambda_{{\rm mfp}i} \gg 1$. 
The cascade of these modes is not fully understood and they will not
be considered here (see \citet{Schekochihin:2007} for further
discussion).

Some fraction of the energy in Alfv\'enic fluctuations at
$k_\perp\rho_i \sim 1$ is expected to launch a kinetic \Alfven wave
cascade to smaller scales \citep{Gruzinov:1998,Quataert:1999}.  In the
limit of anisotropic fluctuations with $k_\parallel\ll k_\perp$, this
cascade, in the wavenumber range $k_\perp \rho_i \gg 1$ and $k_\perp
\rho_e \ll 1$, can again be described in fluid-like terms: the
governing system of equations is known as Electron Reduced MHD (ERMHD)
\citep{Schekochihin:2007}.  In the high-beta limit, it coincides with
the anisotropic limit of the better known but nonrigorous fluid model
called Electron MHD (EMHD) \citep{Kingsep:1990}, where the ions are
assumed motionless and the current is carried entirely by the
electrons.  Assuming locality of interactions and a critical balance
between the linear and nonlinear timescales is again a plausible route
towards a scaling theory of the kinetic \Alfven wave cascade. The
outcome of this argument (explained in more detail in
\secref{sec:model}) is the one-dimensional magnetic energy spectrum
$E_{B}(k_\perp) \propto k_\perp^{-7/3}$ and a predicted wavenumber
anisotropy that scales as $k_\parallel\propto k_\perp^{1/3}$
\citep{Biskamp:1999,Cho:2004,Krishan:2004,Dastgeer:2005,Schekochihin:2007}.
Simulations of EMHD turbulence support these
predictions \citep{Biskamp:1999,Cho:2004,Dastgeer:2005}.

\begin{figure}
\resizebox{\textwidth}{!}{\includegraphics{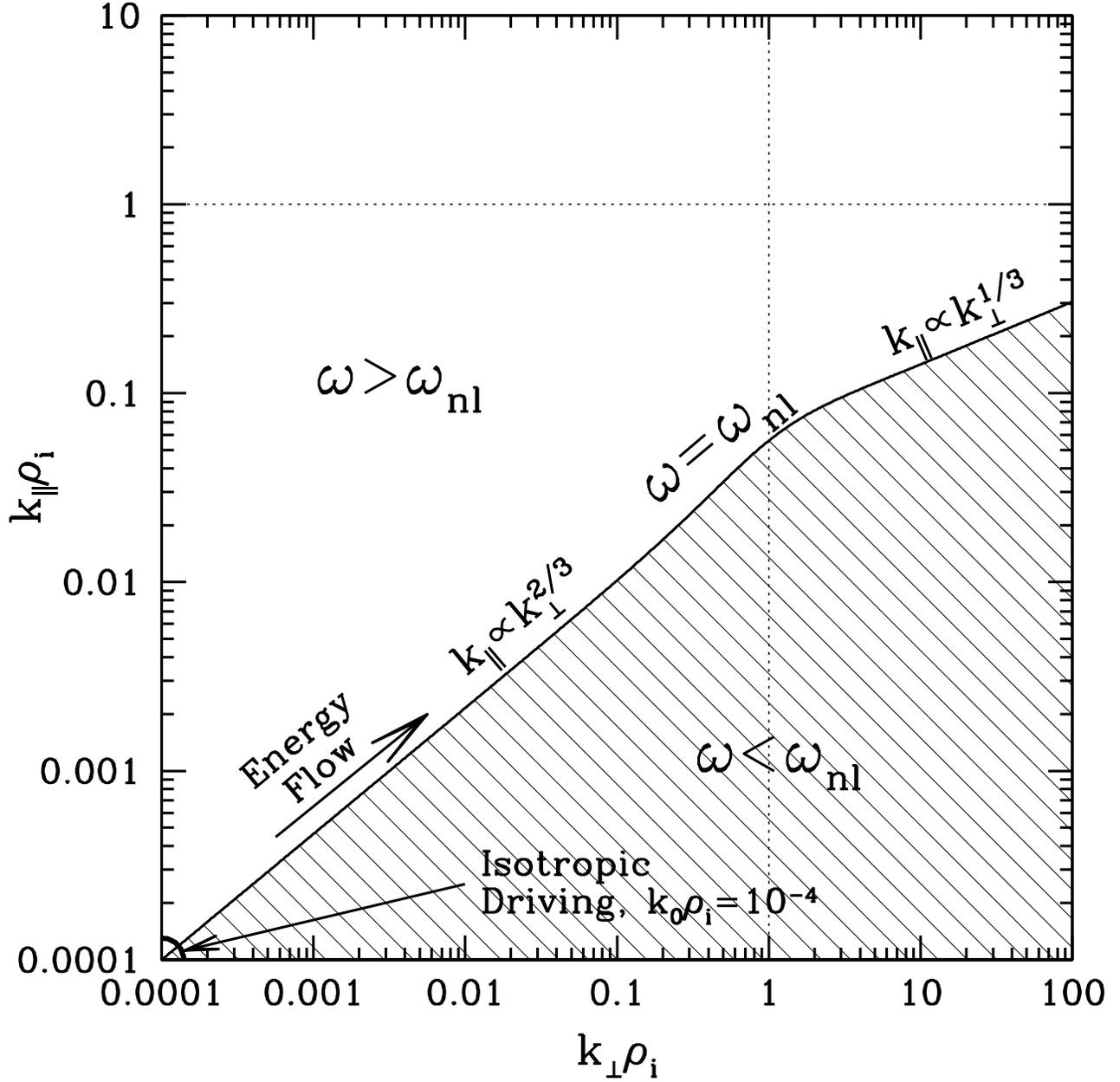}}
\caption{\label{fig:kplane} Schematic diagram of the nonlinear transfer 
of energy across the $(k_\perp,k_\parallel)$ plane from the low
driving wavenumber to high wavenumbers where the turbulence is
dissipated by kinetic processes. Driving is chosen to be isotropic at
a wavenumber $k_0\rho_i=10^{-4}$.  Critical balance,
$\omega=\omega_{nl}$, defines a one-dimensional path for the turbulent
cascade of energy on the $(k_\perp,k_\parallel)$ plane denoted by the
solid line. Numerical simulations show turbulent energy over
the entire shaded region, in which $\omega \le \omega_{nl}$. The cascade model
constructed in this paper can be thought of as describing energy
integrated over all $k_\parallel$ at each value of $k_\perp$.}
\end{figure}

The MHD \Alfven wave and kinetic \Alfven wave cascades correspond to a
nonlinear flow of energy to higher wavenumbers across the
$(k_\perp,k_\parallel)$ plane, as illustrated in \figref{fig:kplane}.
Imposing critical balance constrains the flow of energy to a
one-dimensional path on the plane defined by $\omega=\omega_{nl}$,
delineated on the figure as a solid line. Critical balance
effectively defines a relation between $k_\perp$, $k_\parallel$, and
the fluctuation amplitude $\delta B_\perp$; for strong turbulence
\citep{Goldreich:1995}, choosing values for two of these quantities
specifies the third. The freedom in this relation is conveniently
parameterized by the wavenumber at which the fluctuations are
isotropic, $k_\perp =k_\parallel\equiv k_0$.  In \figref{fig:kplane},
adjusting the value of the isotropic wavenumber $k_0 \rho_i$, or
equivalently adjusting the amplitude of the fluctuations at a given
value of $k_\perp \rho_i$, shifts the curve defining critical
balance vertically; in this figure, we have chosen $k_0 \rho_i=
10^{-4}$, which is the estimated value for the slow solar wind (see
\secref{sec:drivescale}).  Since most treatments of magnetized
turbulence assume an isotropic driving mechanism, we call the
parameter $k_0\rho_i$ the \emph{isotropic driving
wavenumber}.\footnote{Note that, although we parameterize the cascade
in terms of the isotropic driving wavenumber $k_0 \rho_i$, this
parameterization does not require that the cascade be driven
isotropically. A cascade driven both \emph{in critical balance} and
anisotropically, with $k_\perp \ne k_\parallel$, can be characterized
by an effective value of $k_0 \rho_i$ along the extrapolated cascade
path at which $k_\perp=k_\parallel$.}  We note here that critical
balance, $\omega=\omega_{nl}$, is not strictly obeyed in MHD
turbulence. Numerical simulations
\citep{Cho:2000,Maron:2001,Cho:2002,Cho:2003,Oughton:2004} suggest
that the turbulent fluctuation energy does not simply reside along the
path specified by critical balance but approximately fills the shaded region in
\figref{fig:kplane}, where $\omega \lesssim \omega_{nl}$.  The one-dimensional 
cascade model presented in this paper can be thought of as
integrating vertically over all $k_\parallel$ at each value of
$k_\perp$, and then treating that energy as if it were concentrated on
the path of critical balance. 

Although the fluid MHD and EMHD treatments provide important
insights into turbulence in magnetized plasmas, modeling accurately
the nonlinear dynamics, in particular the collisionless damping of
turbulent fluctuations, requires a kinetic treatment.  The inherently
anisotropic nature of magnetized turbulence lends itself to
description by an approximation called gyrokinetics
\citep{Rutherford:1968,Taylor:1968,Catto:1978,Antonsen:1980,Catto:1981,
Frieman:1982,Dubin:1983,Hahm:1988,Brizard:1992,Howes:2006,Brizard:2007,
Schekochihin:2007}. Gyrokinetics is a low-frequency expansion of
kinetic theory valid in the limit of frequencies small compared to the
ion cyclotron frequency $\omega \ll \Omega_i$, ion Larmor radius small
compared to the length scale of the equilibrium quantities $\rho_i \ll
L $, wavevectors of electromagnetic fluctuations nearly
perpendicular to the mean magnetic field $k_\parallel \ll k_\perp$,
and perturbed fields small compared to the mean field $\delta {\bf
B} \ll B_0$. The gyrokinetic equation is derived by averaging over the
particle Larmor motion.  This averaging eliminates the fast wave and
the cyclotron resonances but retains finite Larmor radius effects and
collisionless dissipation via the Landau resonance. The averaging
procedure also eliminates one of the dimensions of velocity in phase
space, reducing the dimensionality of the distribution function from 6
to 5, enabling numerical studies of nonlinear gyrokinetic turbulence
with current computational resources. The validity of the gyrokinetic
description for the solar wind is explored in detail in
\secref{sec:appgk}.

\subsection{Turbulence in the Solar Wind}
\label{sec:solwind}
One of the principal measurements in the study of solar wind
turbulence is the magnetic field fluctuation frequency spectrum
derived from \emph{in situ} satellite observations. At 1~AU, the
one-dimensional energy spectrum in spacecraft-frame frequency
typically shows, for low frequencies, a power law with a slope of $-5/3$,
suggestive of a Kolmogorov-like inertial range
\citep{Goldstein:1995}; a spectral break is typically
observed at around 0.4 Hz, with a steeper power law at higher
frequencies, often denoted the dissipation range in the literature,
with a spectral index that varies from $-2$ to $-4$
\citep{Leamon:1998a,Smith:2006}. The general consensus is that the
$-5/3$ portion of the spectrum is the inertial range of an MHD 
Alfv\'enic turbulent cascade, but the dynamics responsible for the break and steeper
portion of the spectrum is not well understood. Various explanations
for the location of the break in the spectrum have been proposed: that
it is coincident with the ion (proton) cyclotron frequency in the plasma
\citep{Denskat:1983,Goldstein:1994,Leamon:1998a,Gary:1999}, or that
the fluctuation length scale (adopting Taylor's hypothesis to convert
from temporal frequency to spatial wavenumber
\citep{Taylor:1938}) has reached either the ion Larmor radius
\citep{Leamon:1998b,Leamon:1999} or the ion
inertial length \citep{Leamon:2000,Smith:2001b}.  The steepening of
the spectrum at higher wavenumbers has been attributed to ion
cyclotron damping
\citep{Coleman:1968,Denskat:1983,Goldstein:1994,Leamon:1998a,Gary:1999},
Landau damping of kinetic \Alfven waves
\citep{Leamon:1998b,Leamon:1999,Leamon:2000}, or the
dispersive nature of whistler waves
\citep{Stawicki:2001,Krishan:2004,Galtier:2006}.

The energy in turbulent fluctuations in the solar wind is not
distributed isotropically over wavevector space;
\citet{Matthaeus:1990} demonstrated that, at wavenumbers $k_\perp
\rho_i \sim 10^{-3}$, the distribution of energy is anisotropic with
respect to the direction of the mean magnetic field, creating a
``Maltese cross'' pattern on a contour plot of the magnetic field
two-dimensional correlation function. This observation has prompted
attempts to quantify the amount of energy split between nearly
parallel modes (``slab'' modes) and nearly perpendicular modes
(``quasi-2-D'' modes) \citep{Bieber:1996,Leamon:1998a}; results show
that 85--90\% of the power resides in the nearly perpendicular modes.
Rather than an anisotropic MHD turbulent cascade, \citet{Ruderman:1999}
propose that phase mixing could be responsible for the generation of
perpendicular modes.  It was recently found that the two components of
the MHD-scale fluctuations evident in the Maltese cross inhabit
separate environments; \citet{Dasso:2005} decomposed the Maltese cross
into dominantly perpendicular modes within the slow wind and
dominantly parallel modes within the fast wind. It is thus crucial to
distinguish observations in the slow solar wind from those in the
fast.

This is not the only difference between the fast and slow solar
wind that serves as a warning that the dynamics underlying the
turbulence in each system may be distinct. The fast and slow
components of the wind appear to arise from different coronal origins:
the steady fast wind originates from funnels of open fields lines in
coronal holes \citep{Tu:2005}, whereas the unsteady slow wind appears
to come from temporarily open streamers and from active regions of
closed magnetic fields \citep{Habbal:1997,Woo:2004,Woo:2005}.  The
imbalance between anti-sunward and sunward Els\"asser spectra can
reach nearly two orders of magnitude in the fast wind, while the slow
wind has a much smaller imbalance, from a factor of a few to
approximate equality \citep{Grappin:1990,Tu:1990a}. A detailed study
of the electron and ion (proton) temperatures in the solar wind by
\citet{Newbury:1998} revealed a nearly constant mean electron
temperature while the mean ion temperature increased monotonically
with the solar wind speed.  In addition, the sense of the ion
temperature anisotropies differ, with $T_{i\parallel } < T_{i\perp }$
in the fast wind and $T_{i\parallel } > T_{i\perp }$ in the slow
\citep{Kasper:2002,Marsch:2006}. \citet{Horbury:1999} and 
\citet{Marsch:1999} provide reviews of other salient differences 
between the fast and slow solar wind.

\begin{figure}
\resizebox{\textwidth}{!}{\includegraphics{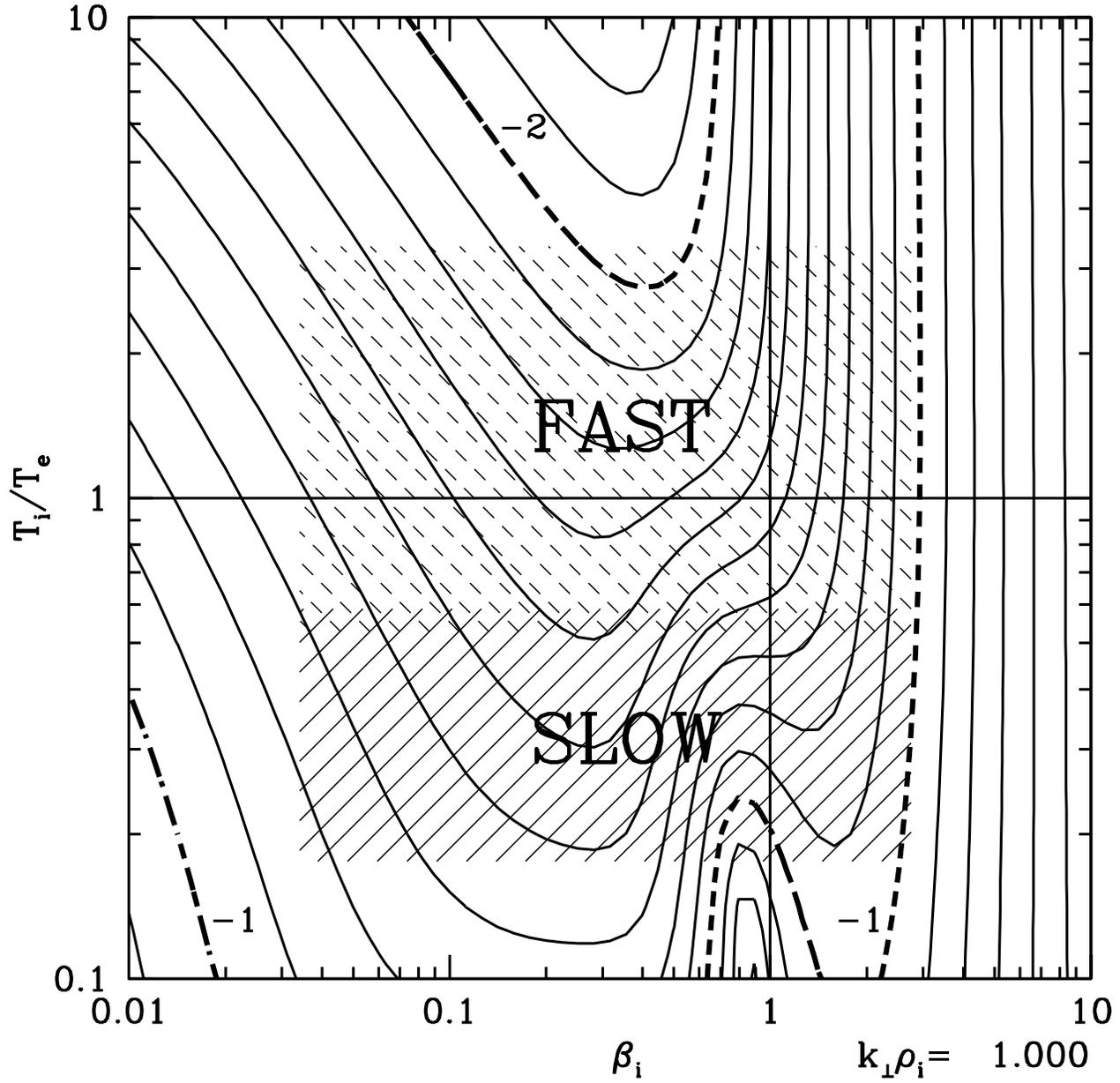}}
\caption{\label{fig:gw} Contours of $\log(\gamma/\omega)$
on the plane $(\beta_i,T_i/T_e)$ for $k_\perp\rho_i =1$. The solid
shading corresponds to the range of parameters in the slow solar wind,
the dashed shading to the fast solar wind. Contours are spaced at
$\Delta \log(\gamma/\omega)=0.1$ with thick dashed lines at each unit
of the logarithm; contour values for each logarithmic unit are
denoted. }
\end{figure}

In this paper, we take heed of the warning by \citet{Tu:1995} of the
dangers of averaging measurements over a long time period because of
the possibility of mixing systems with different properties. A
variation in parameters such as the ion plasma beta $\beta_i$ and the
temperature ratio $T_i/T_e$ can have a strong effect on the kinetic
dynamics of the solar wind, and an imbalance of the sunward and
anti-sunward wave fluxes may strongly affect the nonlinear turbulent
transfer of energy, so the fast streams of solar wind may evolve quite
differently from the slow streams. To provide a concrete example of
the effect on kinetic dynamics, we have calculated the kinetic damping
of linear waves with $k_\perp \rho_i=1$ using the
gyrokinetic\footnote{The Vlasov-Maxwell dispersion relation for a hot
plasma agrees with the gyrokinetic results in the limit $k_\parallel
\ll k_\perp$ and $\omega \ll \Omega_i$, precisely the anisotropic and
low-frequency fluctuations anticipated by the GS theory for MHD
turbulence.}  linear collisionless dispersion relation
\citep{Howes:2006} over the measured range of temperature ratios
$T_i/T_e$ \citep{Newbury:1998} and ion plasma beta $\beta_i$
\citep{Leamon:1998a} (note, however, that the $\beta_i$ range in this paper was 
not sorted by solar wind speed). \figref{fig:gw} presents a contour
plot of $\log(\gamma/\omega)$ on the $(\beta_i,T_i/T_e)$ plane, where
$\omega$ is the linear frequency and $\gamma$ the linear damping rate.
Marked on this plot are the parameter ranges of the slow (solid
shading) and fast (dashed shading) solar wind. Immediately evident is
that the Landau damping\footnote{The Landau resonance includes both
Landau damping by the parallel electric field perturbation and
transit-time, or
\citet{Barnes:1966}, damping by the parallel magnetic field
perturbation; in this paper, we refer to the collective damping effect
of these mechanisms as Landau damping.} is significantly stronger in
the slow solar wind.  It is also clear from \figref{fig:gw} that the
damping rate varies non-trivially with both ion plasma beta $\beta_i$
and temperature ratio $T_i/T_e$.

\subsection{Forward Modeling}
\label{sec:forward}

The complexity of \figref{fig:gw} underscores the challenge of
unraveling the underlying physical mechanisms at work in the
turbulent solar wind from a statistical ensemble of measurements taken
over a wide range of the key parameters: the solar wind speed $V_{\rm sw}$, ion
plasma beta $\beta_i$, and temperature ratio $T_i/T_e$.  Unless two of
these parameters can be held fixed while the third is varied---and
imposing such a constraint would yield a very small number of selected
measurements---extracting meaningful correlations from statistical
analyses of data is very difficult. Compounding this difficulty is the
fact that---although current and planned missions, such as Cluster,
are able to provide simultaneous measurements of the solar wind in
multiple spatial locations---most of our knowledge of the nature of
turbulence in the solar wind has been gleaned through data streams of
single-point measurements in time
\citep{Fredricks:1976,Bruno:2005}. To interpret the data in an attempt
to discern the underlying dynamics of the solar wind, a wide range of
approaches have been employed, many of which rest upon reasonable but
unproven assumptions.  The various approaches to interpreting data
range from studies of the intrinsic variations of the solar wind
parameters as well as their evolution with heliocentric distance, to
detailed examination of the characteristics of fluctuation frequency
spectra and the anisotropic transfer of energy in wavevector space, to
analysis of the fluctuations as linear eigenmodes of the plasma, 
to indirect investigation of turbulent dissipation through
thermodynamic measurements.  Several recent reviews
\citep{Tu:1995,Bruno:2005,Marsch:2006} cover the enormous breadth of
knowledge that has been gained through these methods over decades of
study. All of these methods fall
under the guise of reverse modeling---analyzing observational data in
an attempt to divine the underlying mechanisms governing the evolution
of turbulence in the solar wind.

Even with the wide reaching characterization of the solar wind
that the above methods have achieved, the mechanisms responsible for
the ubiquitous break in the magnetic frequency spectrum between the
inertial and dissipation ranges or for the varying spectral index in
the dissipation range have still not been unequivocally identified
\citep{Smith:2006}. Magnetized turbulence and kinetic plasma physics,
as well as the starkly contrasting behavior of the fast versus the
slow wind, contribute to the overwhelming complexity of the solar
wind, making progress through reverse modeling techniques difficult.
In light of recent advances in the general understanding of turbulence
in magnetized plasmas and of the steady gains in computational
resources, this paper pursues the complementary approach of forward modeling
\citep{Oughton:2005}.

A forward modeling approach has been fruitfully employed to study the
spatial transport and spectral evolution of large-scale, low-frequency
MHD fluctuations in the solar wind \citep{Tu:1984,Tu:1988,Zhou:1990a};
see the review by \citet{Tu:1995} for an extensive discussion of this
well developed field.  In addition, the quest to understand MHD
turbulence has been spearheaded by forward modeling through detailed
numerical simulations \citep{Shebalin:1983,Matthaeus:1996,
Matthaeus:1998,Cho:2000,Muller:2000,Biskamp:2000,Maron:2001,Cho:2002,
Cho:2003,Vestuto:2003,Oughton:2004,Dastgeer:2006}.  To better
understand the turbulent dynamics of the solar wind at frequencies in
the neighborhood of the observed break in the magnetic energy
spectrum, a kinetic treatment is essential
\citep{Marsch:2006,Schekochihin:2007}. 
Yet the complexity of this transition regime also affords the
opportunity to better constrain the underlying physical
mechanisms. Forward modeling through numerical simulation of weakly
collisional turbulent plasmas at this transition has just become
possible with current computational resources and codes based on
gyrokinetic theory. The gyrokinetic framework opens up the opportunity
to study not only the dynamics of the turbulence but also its
dissipation and the consequent particle heating. We believe that,
through detailed nonlinear gyrokinetic simulations of the turbulent
cascade of energy through this regime, the mechanisms responsible for
the break in the magnetic energy spectrum and for the damping of
turbulent fluctuations and particle heating will at last be
identified.

To facilitate contact between gyrokinetic numerical simulations,
necessarily of limited dynamic range, and the wealth of observational
data, we construct a turbulent cascade model. The model combines the
heuristic theories of strong turbulence in the MHD \Alfven and kinetic
\Alfven wave regimes with the precise predictions of linear wave
damping. Although this simple model can account for a number of
observations in the turbulent solar wind, we emphasize the importance
of nonlinear numerical simulations both in justifying the assumptions
made by the model and in enabling detailed comparison to measurements
that go beyond its scope.

\subsection{Outline of the Paper}

In \secref{sec:spectra}, we first review the theory of critically
balanced, magnetized turbulence both in the MHD and kinetic \Alfven
wave regimes, constructing an analytical model that smoothly connects
these regimes (\secref{sec:model}). We then proceed to use the model
to construct magnetic energy spectra consistent with those observed in
the solar wind and to argue that the combination of instrumental
effects and linear damping may be sufficient to explain a wide spread
in the observed spectral indices in the dissipation range. The
validity of gyrokinetics in describing turbulence in the balanced, slow
solar wind, the effective driving scale, the degree of anisotropy,
evidence of a kinetic \Alfven wave cascade from satellite measurements
and the (un)importance of the ion cyclotron resonance are discussed in
\secref{sec:appgk}.  The limitations of our approach and related
previous work are discussed in
\secref{sec:discuss}. Finally, \secref{sec:conc} summarizes 
our findings, paving the path for nonlinear gyrokinetic simulations
of turbulence in the slow solar wind.

\section{Magnetic Energy Spectra in the Slow Solar Wind}
\label{sec:spectra}

Here we construct a model that follows the nonlinear transfer of
energy from the low driving wavenumber, through the inertial range of
the critically balanced MHD \Alfven wave cascade ($k_\perp
\rho_i\ll 1$), connecting to a critically balanced kinetic \Alfven 
wave cascade ($k_\perp \rho_i\gg 1$), and on to the highest
wavenumbers where the turbulent energy is dissipated. To extend the
cascade into the kinetic \Alfven wave regime, we choose to follow the
magnetic energy rather than the kinetic energy of the turbulent
fluctuations because, as the ions decouple from the field fluctuations
at scales below the ion Larmor radius, their kinetic energy becomes
subdominant.  The model assumes that the \Alfven wave energy fluxes
along the magnetic field in either direction are balanced; this
condition, not generally satisfied in the fast solar wind, is
satisfied in at least some intervals of the slow solar wind
\citep{Grappin:1990,Tu:1990a}.  Having set up our model in
\secref{sec:model}, we solve it numerically and present the resulting
spectra in \secref{sec:spec1}--\secref{sec:kc}.  In this section, we
will assume that $\omega \ll \Omega_i$ and that gyrokinetic theory
applies. This assumption will be investigated in detail in
\secref{sec:appgk}, and the limitations of our model are discussed in
\secref{sec:discuss}.

\subsection{Critically Balanced Turbulence and the Turbulent Cascade Model}
\label{sec:model}

Consider a homogeneous magnetized plasma permeated by a mean magnetic field of
magnitude $B_0$ and driven isotropically at an outer scale
wavenumber $k_0$ with velocity $v_0$.
We assume that both the driving scale and the ion mean free path along the magnetic field 
are much larger than the ion Larmor radius, $k_0\rho_i\ll 1$ and $\lambda_{{\rm mfp}i} \gg \rho_i$.

Following Kolmogorov, we adopt the standard assumption that the flux
of turbulent energy (per unit mass and volume) at a given scale is 
entirely determined by the turbulence at that scale 
({\em locality of interactions}). Note that we
make this assumption at {\em all} scales.  We write the energy cascade
rate as
\begin{equation}
\epsilon(\kperp) = C_1^{-3/2} k_\perp v_k b_k^2.
\label{eq:flux}
\end{equation}
The magnetic field fluctuation is written in velocity units, $b_k^2
\equiv \delta B_\perp^2 (k_\perp) /4 \pi n_i m_i$, where $\delta
B_\perp^2(k_\perp)/8 \pi$ is the energy density of the magnetic field
fluctuations perpendicular to the mean field integrated over all
possible values of $k_\parallel$.  The electron fluid velocity
perpendicular to the mean magnetic field is $v_k \equiv
v_\perp(k_\perp)$, which is equal to the ion velocity in the MHD limit,
$k_\perp \rho_i \ll 1$.  All other symbols are defined in
Table~\ref{tab:defs}.  In principle the dimensionless ``{\em
Kolmogorov constant}'' $C_1$ is a function of all ``local''
dimensionless numbers---{\em i.e.} $T_i/T_e$, $\beta_i$,
$k_\perp\rho_i$, $k_\parallel /k_\perp$, 
${\bf v_k}\cdot{\bf b_k}/(v_kb_k)$, 
and $\omega/(k_\perp v_\perp)$, where $\omega$ is the local linear
frequency. In addition to \eqref{eq:flux},  we  assume that the linear and nonlinear time
scales are equal at all scales.  This is the {\em critical
balance} assumption of Goldreich and Sridhar,
\begin{equation}
\omega = \omega_{nl}(\kperp) = C_2 k_\perp v_k.
\label{eq:critbal}
\end{equation}
With this second assumption, the two constants, $C_1$ and $C_2$, depend
on the slightly reduced set of dimensionless numbers: $T_i/T_e$,
$\beta_i$, $k_\perp\rho_i$, $k_\parallel /k_\perp$, and ${\bf v_k}\cdot{\bf
b_k}/(v_kb_k)$.  The critical balance assumption leads, as we have
already explained, to the scale dependent anisotropy, $k_\parallel
\sim k_\parallel(k_\perp) \ll k_\perp$. To make further progress we
must use the dynamical equations. These are reduced MHD (RMHD) in the
\Alfven wave scale range ($\kperp\rho_i\ll1$) and electron reduced MHD
(ERMHD) in the kinetic \Alfven wave range ($\kperp\rho_i\gg1$)---see
\citet{Schekochihin:2007} for a systematic derivation of both sets 
of equations.  Linear theory in both scale ranges relates the magnetic
field and the velocity via
\begin{equation} v_k = \pm \alpha(k_\perp) b_k,
\label{eq:vb}
\end{equation}
where the $\pm$ refers to the direction of wave propagation along
${\bf B}$ and
\begin{equation}
\alpha(k_\perp)   = \left\{ 
\begin{array}{cc} 
1,& k_\perp\rho_i\ll 1  \\
k_\perp \rho_i/
{\sqrt{\beta_i+ 2/(1 + T_e/T_i)}}, & k_\perp\rho_i\gg 1
\end{array} \right.
\label{eq:alpha}
\end{equation}
We shall assume, in the spirit of the critical balance assumption,
that \eqref{eq:vb} holds in the nonlinear cascade.  Similarly, the
linear frequency in gyrokinetic theory is given by
\citep{Howes:2006}
\begin{equation}
\omega = \pm \overline{\omega}(k_\perp) k_\parallel v_A .
\label{eq:omega}
\end{equation}
where $\overline{\omega}(k_\perp) =\alpha(k_\perp)$ in both 
asymptotic ranges $\kperp\rho_i\ll1$ and $\kperp\rho_i\gg1$ 
but not in the transition region $\kperp\rho_i\sim 1$. The
scale invariance of the RMHD and ERMHD equations in their 
respective scale ranges suggests that $C_1$ and $C_2$ are
independent of $k_\perp\rho_i$ and $k_\parallel/k_\perp$ {\em within} 
each scale range. 

The turbulent energy is converted to heat via wave-particle
interactions and collisions.  In the linear regime, this is manifested
as a damping rate, given in the gyrokinetic limit by
\begin{equation}
\gamma = \overline{\gamma}(k_\perp) k_\parallel v_A .
\label{eq:gamma}
\end{equation}
We will assume that the turbulent heating rate per unit mass at each
wavenumber $k_\perp$ is $2\gamma b_k^2$---in other words, that the
linear result persists in the nonlinear regime.  This allows us to
write a one-dimensional continuity equation for the magnetic energy
spectrum in perpendicular wavenumber space, analogous to the standard
simplest spectral evolution model used to describe isotropic
hydrodynamic turbulence \citep{Batchelor:1953},
\begin{equation}
\frac{\partial b_k^2}{\partial t} = 
-k_\perp \frac{\partial \epsilon(k_\perp) }{\partial k_\perp} + S(k_\perp) - 
2{\gamma} b_k^2,
\label{eq:modelcont}
\end{equation}
where the rate of change of magnetic energy at a given perpendicular
wavenumber is given by the three terms on the right-hand side: from
left to right they are the flux of energy in wavenumber space, a
source term representing the driving of the turbulence 
(at a low wavenumber $k_0$), and a damping term
associated with the linear damping rate at a given perpendicular
wavenumber. In steady state, we write
\begin{equation}
 k_\perp \frac{\partial \epsilon(k_\perp) }{\partial k_\perp} = S(k_\perp) - 
2C_1^{3/2}C_2\frac{\overline{\gamma}(k_\perp)}{\overline{\omega}(k_\perp)} 
\epsilon,
\label{eq:modelcont2}
\end{equation}
where we have used equations~(\ref{eq:flux}), (\ref{eq:critbal}),
(\ref{eq:omega}) and (\ref{eq:gamma}) to rewrite the damping
term. Note that because of dissipation the energy cascade rate
$\epsilon$ is not constant but depends on $k_\perp$.  Calculating
$\overline{\gamma}(k_\perp)/\overline{\omega}(k_\perp)$ from linear
gyrokinetic theory \citep{Howes:2006} yields the
\emph{gyrokinetic cascade model}. Equation~(\ref{eq:modelcont2}) can
be integrated if we assume a form for the functions $C_1$ and $C_2$.
Assuming the source $S(k_\perp)$ is localized at $k_0$ we have, for
$k_\perp >k_0$,
\begin{equation}
\epsilon(k_\perp) = \epsilon_0 
\exp{\left\{-\int_{k_0}^{k_\perp}2C_1^{3/2}C_2
\frac{\overline{\gamma}(k_\perp ')}{\overline{\omega}(k_\perp ')}
\frac{dk_\perp '}{k_\perp '}\right\}},
\label{eq:modelcont3}
\end{equation}
where $\epsilon_0$ is the rate of energy input at $k_0$.  

Using equations~(\ref{eq:flux}) and (\ref{eq:vb}),
we obtain the one-dimensional magnetic energy spectrum
\begin{equation}
E_{B}(k_\perp)  = \frac{b_k^2}{k_\perp} = C_{1}\epsilon^{2/3} 
k_\perp^{-5/3}\alpha^{-2/3}.
\label{eq:ekmhd}
\end{equation}
where $\epsilon$ is obtained from evaluating \eqref{eq:modelcont3}.
If dissipation \emph{were} neglected, the energy cascade rate would be
constant $\epsilon=\epsilon_0$ and \eqsref{eq:alpha}{eq:ekmhd} would
lead to the familiar relations: $E_{B}(k_\perp) =
C_{1}\epsilon_0^{2/3} k_\perp^{-5/3}$ in the MHD \Alfven wave regime
and $E_{B}(k_\perp) = C_{1}\epsilon_0^{2/3}
k_\perp^{-7/3}\rho_i^{-2/3}[\beta_i+ 2/(1 + T_e/T_i)]^{1/3}$ in the
kinetic \Alfven wave regime.  We will assume (as before) isotropic driving 
at wavenumber $k_0$.  
Thus assuming equations (\ref{eq:flux}),
(\ref{eq:critbal}), (\ref{eq:vb}), and (\ref{eq:omega}) hold up to the driving scale, 
where $\kpar=\kperp=k_0$ and $k_0\rho_i \ll 1$, 
we obtain $\epsilon_0 = C_1^{-3/2}C_2^{-3}k_0v_A^3$. 

Equations~(\ref{eq:flux}),
(\ref{eq:critbal}), (\ref{eq:vb}), and (\ref{eq:omega}) yield the
typical parallel wavelength at scale $k_\perp$,
\begin{equation}
k_\parallel(\kperp)   = k_\perp^{2/3} k_0^{1/3} \frac{\alpha^{2/3}}{\overline{\omega}} 
\left(\frac{\epsilon}{\epsilon_0}\right)^{1/3}, 
\label{eq:kpara}
\end{equation}
where $\epsilon$ is given by \eqref{eq:modelcont3}.
Note that the normalized frequency is defined by \eqref{eq:omega} and is
calculated from gyrokinetics.  It is sufficient, given the brutality
of our assumptions, to make the approximation $\alpha(\kperp) =
\overline{\omega}(k_\perp)$ throughout the scale ranges. Again, if
dissipation {\em were} neglected, $\epsilon=\epsilon_0$ and
\eqsref{eq:alpha}{eq:kpara} would lead to the familiar relations quoted in 
the introduction: $k_\parallel= k_0^{1/3} k_\perp^{2/3}$ in the MHD
\Alfven wave regime and $k_\parallel =
k_0^{1/3}k_\perp^{1/3}\rho_i^{-1/3}[\beta_i+ 2/(1 + T_e/T_i)]^{1/6}$
in the kinetic \Alfven wave regime.

(Gyro)kinetic theory enters the model we have constructed 
via the computation of the linear frequency $\overline{\omega}(\kperp)$ 
(assuming also $\alpha=\overline{\omega}$) in \eqsref{eq:ekmhd}{eq:kpara} 
and also the linear damping rate $\overline{\gamma}(\kperp)$ 
in \eqref{eq:modelcont3}. The undetermined 
Kolmogorov constants $C_1$ and $C_2$ 
represent a considerable source of uncertainty, especially because 
they may depend on the plasma parameters $\beta_i$ and $T_i/T_e$ 
and, around $\kperp\rho_i\sim1$, also on $\kperp$. While acknowledging 
this uncertainty, we believe it is useful to look at the implications 
of our model with some fiducial values of the constants. 
We note that $C_1$ simply normalizes the spectrum, so we only 
need to specify the product $C_1^{3/2}C_2$ in \eqref{eq:modelcont3}, 
which parameterizes the overall strength of the collisionless damping. 
We use the value $C_1^{3/2}C_2=6$, which was originally inspired 
by the values of the constants measured in MHD simulations of the \Alfven 
wave regime ($C_{1} \simeq 2.5$ and $C_{2} \simeq 1.4$ \citep{Maron:1998pc}; 
these values were used also in \citet{Quataert:1999}). Without similar numerical
guidance in the kinetic regime, we take the constants to be the same
at all scales. While this guess is expedient it is obviously not entirely
satisfactory. Nonetheless, we evaluate \eqref{eq:modelcont3} for
$\epsilon (k_\perp)$ (numerically) with these assumptions. We will
study the sensitivity of our results to the values of $C_1$ and $C_2$ in
\secref{sec:kc}. The values of these constants will ultimately be determined by
future gyrokinetic simulations.


\subsection{A Typical Magnetic Energy Spectrum}
\label{sec:spec1}

Let us now present steady-state magnetic fluctuation energy
spectra from the numerical solution of the gyrokinetic cascade model
constructed in \secref{sec:model}. All spectra are normalized such that
the spectrum value is 1 at $k_\perp = k_{\perp i} \ll \rho_i^{-1}$, the lowest plotted
perpendicular wavenumber. Thus, we plot 
\begin{equation}
\hat{E}_B (k_\perp)=E_B(k_\perp)/E_B(k_{\perp i})
\end{equation}
where $E_B(k_{\perp i}) =(k_0/k_{\perp i})^{2/3} v_A^2/k_{\perp i}
C_2^2$.  Note that, given the assumption $\omega\ll\Omega_i$, the
solution to the model of \secref{sec:model} is formally independent of
the value of the driving wavenumber $k_0 \rho_i$; this parameter only
affects the overall normalization of the spectra plotted here.

\begin{figure}
\resizebox{\textwidth}{!}{\includegraphics{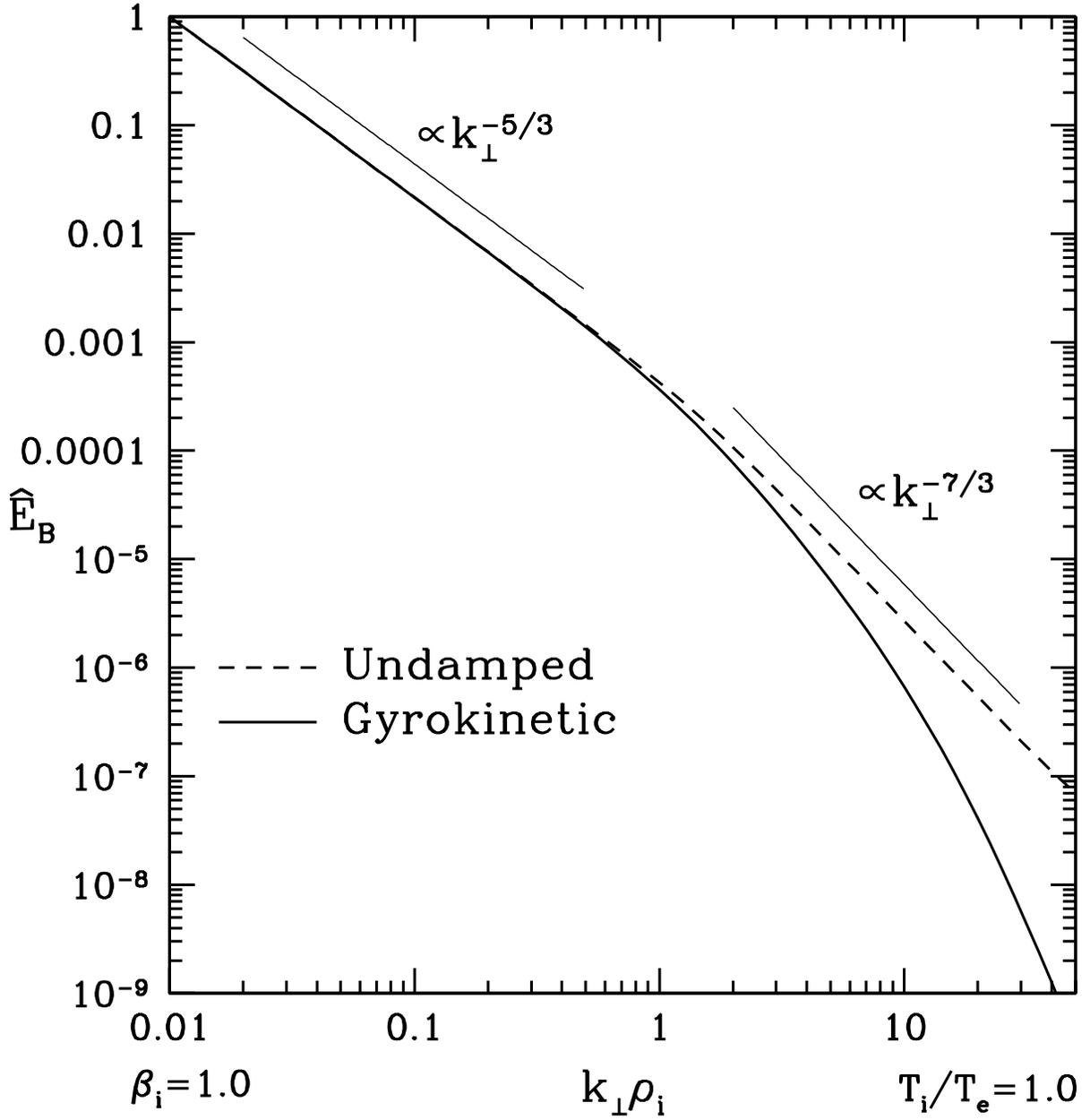}}
\caption{\label{fig:2models}  One-dimensional magnetic energy
spectra for the turbulent cascade from the MHD to the
kinetic \Alfven wave regime. The chosen parameters are $\beta_i=1$ and
$T_i/T_e=1$. Solutions of our cascade model with no damping (dashed) and
with the gyrokinetic damping rate (solid) are shown.}
\end{figure}

In \figref{fig:2models} we plot the normalized one-dimensional magnetic fluctuation 
energy spectrum, $\hat{E}_B(k_\perp)$, as a function of wavenumber
$k_\perp \rho_i$ for a plasma with ion plasma beta $\beta_i=1$ and
temperature ratio $T_i/T_e=1$.  A plasma of protons and electrons is
assumed for all the steady-state spectra presented in this paper. Two
cases are shown: the solution of the gyrokinetic cascade model of
\secref{sec:model} using the appropriate gyrokinetic damping rate
$\overline{\gamma}(k_\perp)$, and, for comparison, an undamped case 
with the damping term artificially set to zero,
$\overline{\gamma}(k_\perp)=0$.  The undamped model recovers the
spectral indices of $-5/3$ in the MHD regime and $-7/3$ in the kinetic
\Alfven wave regime, as predicted by the arguments neglecting
dissipation in \secref{sec:model}.  The damping at $k_\perp \rho_i
\gtrsim 1$ in the gyrokinetic model is sufficient to cause the
spectrum to fall off more steeply than the undamped prediction of
$-7/3$.  The steady-state gyrokinetic spectrum obtained here clearly
demonstrates the exponential roll-off characteristic of dissipation
\citep{Li:2001}.

\subsection{Effect of Limited Magnetometer Sensitivity}
\label{sec:noise}
The magnetic fluctuation spectrum at frequencies above the spectral
break from \emph{in situ} solar wind measurements is widely
interpreted to behave like a power law rather than an exponential
decay.  We suggest here that the power-law appearance of the spectrum
in this range may be an instrumental effect.  The power spectrum of
magnetic fluctuations from satellite measurements is limited by the
fluxgate magnetometer sensitivity; this sensitivity limit can, e.g.,
be clearly seen in Figure~6 of \citet{Leamon:1998a} at the high
frequency end of the spectrum. The noise floor of a fluxgate
magnetometer at frequencies $f> 1$ Hz is constant in units of
nT/$\sqrt{\mbox{Hz}}$ \citep{Bale:2007pc}; hence, on a plot of the
one-dimensional magnetic energy spectrum, this corresponds to a
constant noise floor at the high frequency end. 

\begin{figure}
\resizebox{\textwidth}{!}{\includegraphics{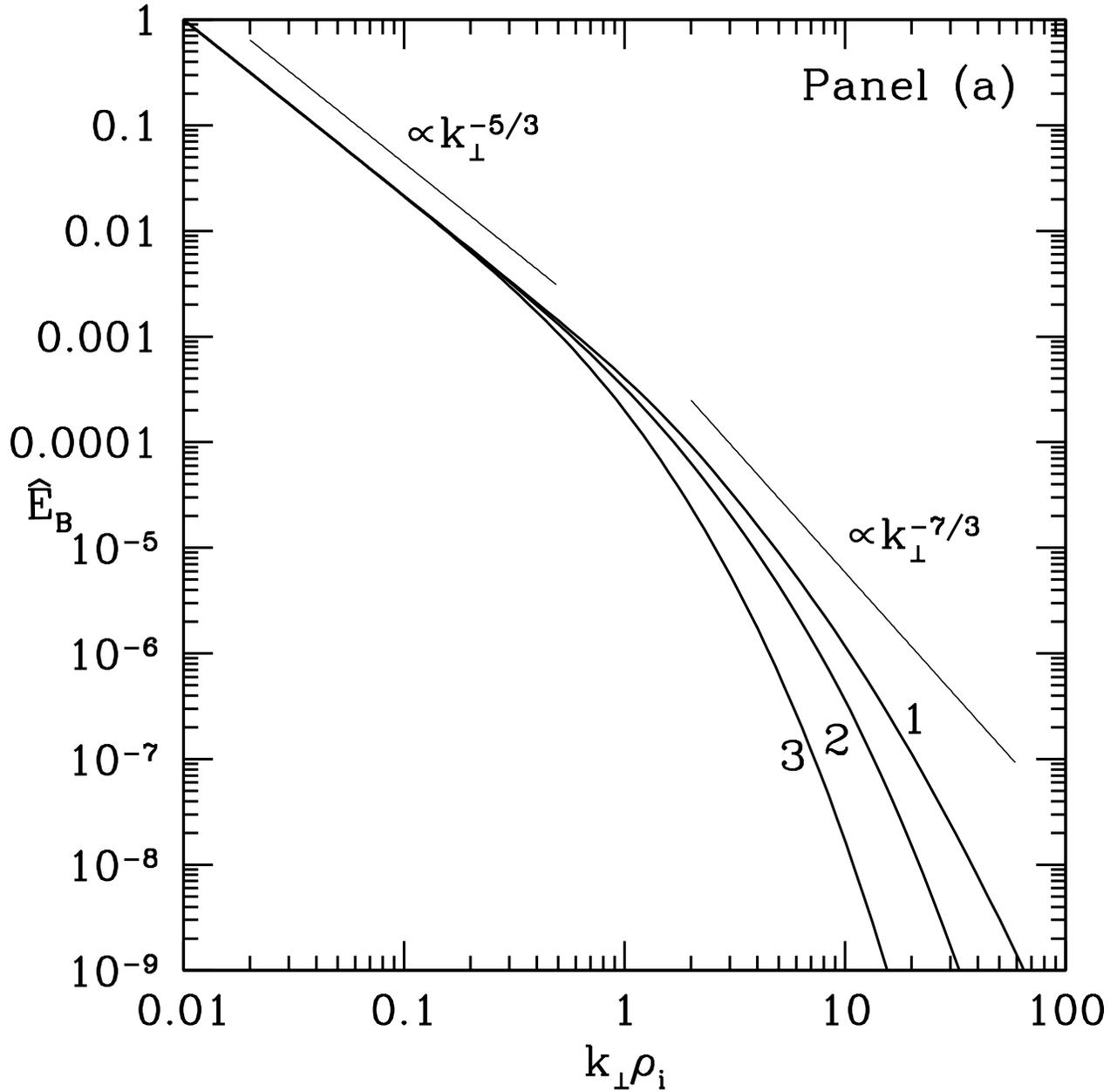}}
\caption{\label{fig:diss_range_a} One-dimensional magnetic energy
spectra for three gyrokinetic models: (1) $\beta_i=0.5$, $T_i/T_e=3$,
(2) $\beta_i=3$, $T_i/T_e=0.6$, (3) $\beta_i=0.03$, $T_i/T_e=0.175$.
Panel (a) shows that all three spectra demonstrate a dissipative
roll-off that cannot be fit by a single spectral index
$k_\perp\rho_i>1$ (consistent with an exponential roll-off). }
\end{figure}

To test the effect of the fluxgate magnetometer noise floor on the
appearance of the magnetic fluctuation spectrum from our gyrokinetic
cascade model, we mock up the instrumental noise by specifying a
constant background value to the one-dimensional energy spectrum.
Inspired by the spectra in Figure~6 of \citet{Leamon:1998a}, we choose
this noise floor to be approximately two to three orders of magnitude
below the spectrum value at the $\kperp\rho_i=1$ break.  In
\figref{fig:diss_range_a} we plot the spectra for three steady-state
solutions of our model with the following parameters, chosen to sample
across the range of solar wind parameters depicted in
\figref{fig:gw}: (1) $\beta_i=0.5$, $T_i/T_e=3$; (2) $\beta_i=3$,
$T_i/T_e=0.6$; and (3) $\beta_i=0.03$, $T_i/T_e=0.175$. All cases
include the damping rate calculated gyrokinetically. 
Panel (a) shows the spectra without the added
constant sensitivity limit; panel (b) adds a constant sensitivity
level at two (spectrum 1) or three (spectra 2 and 3) orders of
magnitude below the spectrum value at $k_\perp \rho_i =1$. The
behavior of each spectrum in panel (b) in the range $k_\perp
\rho_i >1$ more closely resembles a power law than the exponential
roll-off, which would characterize the noiseless spectra; the
steady-state solutions are well-fit by power laws with spectral
indices $-7/3$, $-3$, and $-4$. In summary, the instrumental
sensitivity limit may produce spectra that appear to obey a power-law
scaling with some effective spectral index even though the underlying
spectrum actually has an exponential roll-off.

At present there exists little evidence in the literature to test this
prediction. Search coil magnetometers are more sensitive at these high
frequencies. \citet{Denskat:1983} studied the magnetic field power
spectra up to 470~Hz using data from both fluxgate and search coil
magnetometers. The search coil data shows power-law behavior up to
470~Hz, but the amplitude of the spectral density drops about three
orders of magnitude in the gap from the highest frequency fluxgate
measurement at 2~Hz to the lowest frequency search coil measurement at
4.7~Hz, suggesting that the higher frequency power law is not a
continuation of the lower frequency spectrum. Alternatively, the
mismatch may be due to difficulty performing a relative calibration of
the instruments, but without an overlap in frequency the difference
cannot be resolved. One recent paper \citep{Alexandrova:2008} shows
search coil measurements of magnetic field spectra up to $10$~Hz, a
factor of 20 in frequency above the breakpoint; for the 57-minute
interval of the fast solar wind studied, the dissipation range
spectrum shows neither an exponential roll-off nor evidence of
instrumental noise up to 10~Hz. During one 7-minute interval, the
dissipation range index is $-2.33$; during another 7-minute interval,
it is $-2.50$; and over the whole 57-minute interval, it is $-2.6$.
These results are not at odds with our model predictions for weak
damping---the dissipation range spectrum is expected to appear as a
$-7/3$ power law to higher frequencies before the damping becomes
strong enough to cause the spectrum to roll off exponentially. We must
also keep in mind that the fast wind generally does not have a balance
of Els\"asser energy fluxes, an assumption of the present simple
cascade model. An interesting aspect of the measurements in
\citet{Alexandrova:2008} not discussed in their paper is the time
variability of the spectral index in the dissipation range. There is a
clear need for a detailed study of solar wind turbulence at
frequencies above 1~Hz, taking into account the varying plasma
parameters, the imbalance in the Els\"asser energy fluxes, the
sensitivity limits of the various instruments, and the time
variability of the dissipation range spectral index.

It is instructive to consider what we will learn if such a study does
not find magnetic energy spectra demonstrating a slow, exponential
roll-off.  The absence of an exponential decay could indicate that 
using the linear damping rates in a turbulent plasma is incorrect.
It is important to emphasize that the predictions of gyrokinetic
theory for solar wind turbulence are not equivalent to the predictions
of this simple cascade model---nonlinear gyrokinetic simulations of
turbulence will play a key role in testing this model, assessing the
robustness of the predicted exponential fall-off, and refining the
model as necessary.

\subsection{The Effective Spectral Index in the Dissipation Range}
\label{sec:index}

\figref{fig:diss_range_a} shows that, over the range of the plasma 
parameters $\beta_i$ and $T_i/T_e$ measured in the solar wind, the
observed wide spread of the spectral index in the dissipation range
from $-2$ to $-4$
\citep{Leamon:1998a,Smith:2006} can be explained by the
damping of kinetic \Alfven waves via the Landau resonance. If the
Landau damping is negligible, the spectral index is expected to be
$-7/3$, close to the observed upper limit; over the range of
parameters relevant to the solar wind, our cascade model gives a lower
limit to the (effective) spectral index of about $-4$, consistent with
observations. The measured effective spectral index for $k_\perp
\rho_i >1$ is a complicated function of the parameters $\beta_i$,
$T_i/T_e$, and $k_0 \rho_i$ as well as of the sensitivity limit of the
magnetometer. If it is indeed the linear damping that explains the
wide spread in the measured values, the complexity of this parameter
dependence is a natural outcome of the nontrivial behavior of the
collisionless damping as a function of $\beta_i$ and $T_i/T_e$ over
the range of parameters in the slow and fast solar wind depicted in
\figref{fig:gw}. Plots of the measured spectral index versus a single
plasma parameter are unlikely to reveal strong correlations due to
variation in the other parameters within the data set. To make headway
here, a forward modeling approach may prove more fruitful. One needs
to incorporate all of the measured parameters to determine the
predicted spectral index for an interval of solar wind data.

\subsection{Uncertainty in the Kolmogorov Constants}
\label{sec:kc}
The spectrum derived from our cascade model depends on the values of
the Kolmogorov constants in the theory. As we noted above, while $C_1$ 
merely normalizes the spectrum, the product $C_1^{3/2}C_2$ affects 
the competition of the nonlinear transfer of energy
with the linear damping [see \eqref{eq:modelcont3}]. 
If this product is increased (decreased), this effectively
makes the linear dissipation stronger (weaker) relative to the
nonlinear transfer. These order unity constants must be determined
from nonlinear simulations or experimental data.  
To judge the effect of varying these constants on the steady-state solution,
\figref{fig:varykc} presents the gyrokinetic spectrum for $\beta_i=1$
and $T_i/T_e=1$, with this product taken to be
$C_1^{3/2}C_2=3,6,\mbox{ and }12$.  The resulting variation in the
effective spectral index is $+1/3$ or $-2/3$, so all of our steady-state spectra 
suffer this systematic uncertainty. In the rest of this paper we use the 
fiducial value $C_1^{3/2}C_2=6$. 

\section{Applicability of Gyrokinetics to the Slow Solar Wind}
\label{sec:appgk}

In this section, we argue that gyrokinetics is applicable to the
dynamics of the turbulence in the balanced, slow solar wind at the
dissipation scale.  Gyrokinetics is \emph{not}, of course, applicable
at the driving scale $L=2\pi/k_0$, where the magnetic field
fluctuations are of the order of the mean field and are relatively
isotropic.  Formally, as described in \secref{sec:turb}, a number of
conditions must
\emph{all} be met for the gyrokinetic approach to be valid: $ \rho_i \ll L$, 
$k_\parallel \ll k_\perp$, $\delta \V{B} \ll B_0$, and $\omega \ll \Omega_i$. 
Let us consider whether each of these conditions is satisfied in 
the solar wind. In \secref{sec:drivescale} we estimate the 
effective driving scale ($ \rho_i \ll L$). An estimate of  
the anisotropy $k_\parallel \ll k_\perp$ at the dissipation scale 
$k_\perp \rho_i \sim 1$ follows in \secref{sec:aniso}. 
The observed amplitude of the fluctuations $\delta \V{B} \ll B_0$ 
is discussed in \secref{sec:amplitude}.  
In \secref{sec:phasevel} we present wave phase velocity evidence from 
satellite measurements that is consistent with a kinetic
\Alfven wave cascade but inconsistent with the onset of ion cyclotron
damping---indirect evidence for $\omega \ll \Omega_i$. In 
\secref{sec:cyclo} we show (in some detail) that the turbulent frequencies 
are usually less than the cyclotron frequency ($\omega \ll
\Omega_i$). Finally, in \secref{sec:conversion}, we explain what
happens if the cyclotron resonance is reached in the dissipation
range.

\subsection{Turbulence Driving Scale}
\label{sec:drivescale}
The nature of the turbulent fluctuations in the dissipation range of
the solar wind depends on the driving scale $L$ at which energy is
injected into the turbulent cascade. The anisotropy predicted by the
critically balanced \Alfven wave cascade in GS theory is scale
dependent, so this raises two questions: is the solar wind driven
isotropically or anisotropically, and what is the effective driving
scale $L$?  Unfortunately, the mechanisms responsible for driving
turbulence in the solar wind are unknown; the matter is further
complicated by the radial expansion of the solar wind as it travels
outward from the sun and the measured imbalance of anti-sunward
vs.~sunward directed \Alfven wave flux
\citep{Grappin:1990,Tu:1990a,Tu:1990b}. Nevertheless, 
we can make some headway on this complex problem if we ignore the
specific details of the driving mechanism and the radial expansion and
simply estimate the anisotropy of the balanced, slow solar wind
turbulence from \emph{in situ} measurements at 1~AU.

As discussed in \secref{sec:turb}, determining the isotropic driving
wavenumber $k_0 \rho_i=2\pi/L$ fixes the path of the critically
balanced cascade as depicted in \figref{fig:kplane}. Equivalently, a
measurement of the wavenumber anisotropy at any scale in the solar
wind can also be used to fix the critical balance relation; here we
use this measurement to estimate the resulting effective value of $k_0
\rho_i$.  From measurements of the anisotropy of the spatial
correlation of magnetic field fluctuations in the slow solar wind,
presented in the left panel of Figure~1 from
\citet{Dasso:2005}, we estimate the anisotropy $k_\perp/k_\parallel =
r_\parallel/r_\perp \simeq 1.5$ at an average scale $2 \pi/k_\perp = 6
\times 10^{10}$~cm.  If we assume that the turbulence is strong at the
isotropic driving scale and that the energy cascade rate is a constant
$\epsilon=\epsilon_0$ between the driving and the observed scales (a
reasonable assumption given that observed spectra consistently show a
spectral index of $-5/3$
\citep{Smith:2006} and that the damping rate for \Alfven waves  
is small, $\gamma \ll \omega$, at these scales), we can estimate
the driving scale from which the observed turbulence arose using
\eqref{eq:kpara},
\begin{equation}
 k_0\rho_i = \left(\frac{k_\parallel}{k_\perp}\right)^3  k_\perp \rho_i.
\label{eq:k0}
\end{equation}
which applies for scales in the MHD \Alfven wave regime $k_\perp \rho_i \ll
1$.  Using a mean ion (proton) temperature for the slow wind $T_i \simeq 4.5
\times 10^4$~K \citep{Newbury:1998} and the geometric mean of the
range of magnetic field magnitude in \citet{Leamon:1998a} of $B_0
\simeq 10^{-4}$~G, we obtain the ion Larmor radius 
$\rho_i \simeq 3 \times 10^6$ cm. Hence, we find $k_0 \rho_i\simeq  10^{-4}$, 
or an isotropic driving scale $L \simeq 2\times 10^{11}$~cm.  

This value is consistent with the estimate by \citet{Matthaeus:2005}
of the mean magnetic field correlation length in the solar wind (over
all solar wind speeds) of $L \simeq 1.2 \times 10^{11}$~cm. Figure~3
of \citet{Tu:1990b} presents spectra of the outward directed
Els\"asser flux of \Alfven waves measured by the Helios spacecraft
missions which show, for the slow solar wind at 0.9~AU,\footnote{There
is a typo in the legend of this figure: the spectrum at 0.9~AU is for
the slow solar wind (L), not the fast (H).} the beginning of the
$-5/3$ spectrum at a reduced wavenumber $k/2 \pi
\simeq 10^{-7}\mbox{ km}^{-1}$, corresponding to a driving scale of $L=
10^{12}$~cm. Finally, an estimate of the correlation length of the
anti-sunward Els\"asser field at 0.3~AU is $L=1.6 \times 10^{11}$ cm
\citep{Tu:1995}; scaling this length for radial expansion to 1~AU
gives a value of $L=5 \times 10^{11}$ cm. These four independent
determinations of the driving scale, at which we assume the
fluctuations are isotropic, are consistent to within an order of
magnitude.  This suggests that the driving is at least
approximately isotropic and that $k_0 \rho_i = 10^{-4}$ is a good
choice for a fiducial value of the isotropic driving wavenumber.

It is perhaps appropriate to discuss why this number is physically
sensible. The simplest interpretation of solar wind turbulence is that
energy resides in large-scale, energy containing modes driven in the
solar corona, and those modes are subsequently left to decay freely in
the wind within the inner heliosphere. A reasonable estimate for the
decay time of turbulent energy characterized by a velocity $v_\perp$
at scale $k_\perp$ is the eddy turnover time $\tau \sim (k_\perp
v_\perp)^{-1}$. For the driving scale estimated above, $L=2
\pi/k_\perp =2\times 10^{11}$~cm, and a turbulent velocity estimated
from the slow solar wind Els\"asser spectra in Figure~4 of
\citet{Grappin:1990} of $v_\perp = 2 $~km/s, this decay time is $\tau
\sim 50$~h. The travel time for slow solar wind with a velocity
$V_{\rm sw} \sim 350$~km/s to reach $R=1$~AU is $t=R/v_{sw} \sim
100$~h. If there is energy in much larger scales than our estimated
value for $L$, those larger scale fluctuations will not have had time
to transfer their energy in the nonlinear cascade.  On the other hand,
energy in modes much smaller than $L$ in the solar corona would have
already cascaded to the dissipation scale and would not be observed at
1~AU. Indeed the near agreement between the estimated turbulent decay
time, $\tau \sim 50$~h, and the travel time of the slow solar wind to
1~AU, $t \sim 100$~h, is probably not coincidental. It suggests that,
within 1~AU, the turbulence in the slow solar wind is freely decaying
and that we are seeing the energy injected at the largest scale that
can cascade on the travel time.


\subsection{Anisotropy at the Ion Larmor Radius Scale}
\label{sec:aniso}

Assuming an ion number density in the slow solar wind of $n_i
\simeq 20$~cm$^{-3}$ \citet{Leamon:1998a} and an electron 
temperature of $T_e \simeq 1.3 \times 10^5$~K \citep{Newbury:1998}, we
find the ion plasma beta $\beta_i \simeq 0.3$ and temperature ratio
$T_i/T_e\simeq 0.35$; for these parameters, the normalized frequency
from the gyrokinetic dispersion relation at $k_\perp
\rho_i=1$ is $\overline{\omega} \simeq 1.45$.  
Substituting these values, along with our fiducial isotropic driving
wavenumber $k_0 \rho_i = 10^{-4}$, into \eqref{eq:kpara} and taking
for simplicity $\epsilon=\epsilon_0$, we obtain the estimate
$k_\parallel /k_\perp
\simeq 0.04$ at $\kperp\rho_i=1$. As will be demonstrated in
\figref{fig:undamped}, by neglecting dissipation, this estimate
represents an upper limit on the value of $k_\parallel/k_\perp$.

\subsection{Fluctuation Amplitude in the Dissipation Range}
\label{sec:amplitude}

One might be concerned that gyrokinetics requires a strong mean
magnetic field in order for the gyroaveraging procedure to be valid,
and that the solar wind does not contain such a strong mean 
field.  At the driving scale $L$ of the turbulence in the
solar wind, the magnetic field has fluctuations of the same order as
the mean field, $\delta \V{B} \sim B_0$. However, as was 
realized by \citet{Kraichnan:1965}, even in a plasma with no
mean magnetic field, the magnetic field of the large-scale
fluctuations can serve as an effective mean field for smaller-amplitude
fluctuations on smaller scales. Raw magnetometer
data, for instance Figure~3-5a of \citet{Tu:1995} or Figure~1 of
\citet{Bale:2005}, show order one
fluctuations of the magnetic field vector components, corresponding to
large changes in the direction of the mean magnetic field (the
magnitude of the field has much smaller fractional fluctuations) on
timescales from 15 minutes to 1 hour. On shorter timescales, however,
the field direction remains relatively constant with small amplitude,
short-timescale fluctuations about the mean. This is in agreement with
Kraichnan's hypothesis, showing fluctuations of both small amplitude
and scale about a mean magnetic field arising from the larger scale
fluctuations---these small fluctuations are precisely those which are
well-described by gyrokinetics. Once the cascade has progressed to
scales a couple of orders of magnitude smaller than the driving scale,
the amplitude of the fluctuations becomes small compared to the local
mean field amplitude and the fluctuations become significantly
anisotropic, $\delta \V{B}/B_0\ll 1$ and $k_\parallel \ll k_\perp$.
Gyrokinetics then becomes a valid description of the fluctuations, and
the deeper one goes into the inertial range, the better it works.
From \eqref{eq:ekmhd}, taking $\epsilon=\epsilon_0$, we have
\begin{equation}
\frac{|\delta \V{B}|_k}{B_0} = {b_k\over v_A} = 
\frac{1}{C_2}\lt(\frac{k_0}{k_\perp}\rt)^{1/3}\alpha^{-1/3}.
\label{eq:deltaB}
\end{equation}
Thus, for our estimated $k_0$ we have $\delta \V{B}/B_0\lesssim 0.03$ 
for $\kperp\rho_i\gtrsim1$.

\subsection{Evidence against the Cyclotron Resonance}
\label{sec:phasevel}
There exists evidence from \emph{in situ} satellite measurements
consistent with the existence of a kinetic \Alfven wave cascade, at
least in some cases.  Wave phase velocity information in the region of
the spectral break can be used to distinguish the characteristics of
the underlying wave modes. \citet{Bale:2005} have produced the first
simultaneous measurements of the fluctuating electric and magnetic
fields, demonstrating that the wave phase velocity above the spectral
breakpoint increases, as shown in panel (b) of Figure~3 in
\citet{Bale:2005}. Using the Taylor hypothesis to convert the
frequency spectrum to a wavenumber spectrum, the spectral breakpoint
occurs near the ion Larmor radius, $k \rho_i \sim 1$.

Using the full Vlasov-Maxwell linear dispersion relation for a hot
collisionless plasma (see \citet{Quataert:1998} for a description of
the code that we use to solve it), we plot in \figref{fig:w_kpar} the
normalized wave phase velocity $\omega/k_\parallel v_A$ vs.~the
normalized wavenumber $k \rho_i$. Plasma parameters are chosen to be
$\beta_i = 1$ and $T_i/T_e = 1$. Although in the gyrokinetic theory
$\omega/k_\parallel v_A =\overline{\omega}(k_\perp) $ is a function of
$k_\perp$ alone, in hot plasma theory $\omega/k_\parallel v_A$ depends
also on $k_\parallel$.  The parallel wavenumber needed for this
calculation is determined by using \eqref{eq:kpara} with no
dissipation, $\epsilon = \epsilon_0$.  The change in wave phase
velocity is plotted for driving scales $k_0 \rho_i = 10^{-6}$,
$10^{-4}$, $10^{-2}$, and $1$.  The wave phase velocity increases
until the ion cyclotron frequency is reached; then, the frequency
asymptotes $\omega\to\Omega_i$ as ion cyclotron damping becomes
important (see Appendix \ref{app:cycdamping}), so $\omega/\kpar v_A$
decreases due to the increasing $\kpar(\kperp)$ [see
\eqref{eq:kpara}].  The measured wave phase velocity in Figure~3 of
\citet{Bale:2005} does not show this downturn, suggesting that the ion
cyclotron damping remains unimportant below the highest wavenumber
measured, $k_\perp \rho_i \simeq 10$. In the limit $k_\parallel \ll
k_\perp$, the upturn of the wave phase velocity above the breakpoint
in the magnetic fluctuation spectrum is consistent with underlying
wave modes consisting of kinetic \Alfven waves with
$\omega\ll\Omega_i$. We do note, however, that this increase of phase
velocity may also be consistent with fast MHD/whistler mode of
Vlasov-Maxwell kinetic theory for certain wavevector anisotropies. For
either case, though, it appears that the break is not due to the onset
of ion cyclotron damping.

\subsection{Conditions for Reaching the Ion Cyclotron Resonance}
\label{sec:cyclo}

Although the \citet{Bale:2005} measurements of the electric fields in
the solar wind are consistent with kinetic \Alfven waves with
frequencies $\omega\ll\Omega_i$, there is also indirect evidence that
fluctuations at or above the cyclotron frequency are present in the
solar corona and solar wind. For example, minor ions such as 0$^{+5}$
are selectively heated to high perpendicular temperatures
\citep{Kohl:1997,Kohl:1998}. Also, the adiabatic invariant $\mu
\propto T_\perp/B$ for ions and alpha particles is observed to
increase with heliocentric distance
\citep{Marsch:1983}, indicating some mechanism for perpendicular heating
of these species, particularly in the fast wind. Since (the
collisionless part of) gyrokinetics preserves the adiabatic invariant
$\mu$, it cannot describe such heating.  This heating may come from
fast wave fluctuations \citep{Chandran:2006} or velocity-space
instabilities
\citep{Scarf:1967,Eviatar:1970,Gary:1976,Kasper:2002,Hellinger:2006,Marsch:2006} rather than the \Alfven
wave cascade, but it is nevertheless important to know where in
$(\kperp,\kpar)$ space the turbulent cascade may start to be affected
by the cyclotron resonance.  In this section we employ the
steady-state solutions of the turbulent cascade model presented in
\secref{sec:spectra} to estimate the perpendicular
wavenumber at which the cascade reaches the cyclotron frequency---this
is the same wavenumber at which a gyrokinetic treatment fails to be
valid.  

From \eqref{eq:omega} and \eqref{eq:kpara} we obtain
\begin{eqnarray}
\frac{\omega}{\Omega_i} = 
\frac{\kpar \rho_i}{\sqrt{\beta_i}}\overline{\omega}(k_\perp) 
= (k_\perp \rho_i)^{2/3}(k_{0} \rho_i)^{1/3}\frac{\overline{\omega}^{2/3}}{\sqrt{\beta_i}}
\left(\frac{\epsilon}{\epsilon_0}\right)^{1/3}\!\!\!,
\label{eq:womegai2}
\end{eqnarray}
where $\epsilon/\epsilon_0$ is given by \eqref{eq:modelcont3}
and we have assumed $\alpha=\overline{\omega}(\kperp)$ for all $\kperp$.
In \figref{fig:undamped}, we plot $\omega/\Omega_i$ and $k_\parallel
\rho_i$ for both the undamped and gyrokinetic cascade models with
$\beta_i=1$, $T_i/T_e=1$, and $k_0 \rho_i=10^{-4}$. It is unknown if
the linear damping rates are a good estimate of the damping of
nonlinear turbulent fluctuations, but the undamped model certainly
provides upper limits for the frequency and parallel wavenumber.
\figref{fig:undamped} shows that the undamped model reaches the
cyclotron frequency at $k_\perp
\rho_i \simeq 12$, whereas the model using the linear gyrokinetic
damping rates never reaches the cyclotron frequency.  The plot of
$k_\parallel\rho_i$ vs.~$k_\perp \rho_i$ for the gyrokinetic model in
\figref{fig:undamped} reaches a maximum at $k_\perp \rho_i\simeq 5$
and then $k_\parallel\rho_i$ begins decreasing.  In this typical case,
it is clear that, \emph{if the cyclotron frequency is reached}, it is
only after the fluctuations have already cascaded to a high
perpendicular wavenumber, $k_\perp \rho_i \gg 1$, at which point most
of the energy in the cascade has already been
dissipated. Consequently, the gyrokinetic treatment will be valid over
a large part, if not all, of the dissipation range.

The decrease in $k_\parallel \rho_i$ seen in the solution of the
gyrokinetic model in \figref{fig:undamped} is likely an unphysical
result arising from the imposition of critical balance, a condition
requiring that the turbulence remains strong. The kinetic damping of
the turbulent fluctuations becomes significant at this point, so the
nonlinear frequency begins to fall below the linear
frequency. Instead of $k_\parallel\rho_i$ decreasing, it is
more likely that the turbulence becomes weak, a condition not treated
within our simple cascade model. The realm of weak turbulence,
$\omega>
\omega_{nl}$, exists above the line of critical balance as depicted in
\figref{fig:kplane}.  In weak turbulence theory, the cascade to higher
parallel wavenumber is suppressed
\citep{Goldreich:1997,Ng:1997,Galtier:2000,Lithwick:2003}, so a more
physical result is that the value of $k_\parallel\rho_i$ should remain constant,
rather than decrease, as the cascade progresses to higher $k_\perp
\rho_i$.  Adjusting the model to prevent the value of
$k_\parallel\rho_i$ from decreasing results in a sharper cutoff in the
magnetic energy spectrum at the high wavenumbers where the parallel
wavenumber is held constant---this is the result of a decrease in the
ratio of the nonlinear transfer frequency to the linear damping rate
as the turbulence weakens. The simplicity of our cascade model limits
our ability to determine the correct physical behavior in this regime;
nonlinear simulations are necessary to determine accurately the
behavior of the turbulent cascade as the kinetic damping becomes
significant.

Analogous to our gyrokinetic cascade model, we can construct a
\emph{hot plasma cascade model} using the frequencies and damping
rates from the linear hot plasma dispersion relation
\citep{Stix:1992,Quataert:1998} instead of the gyrokinetic values. 
In hot plasma theory, however, the normalized wave frequency
$\omega/(k_\parallel v_A)$ is no longer a function of
$k_\perp$ alone, but depends also on $k_\parallel$. Since the linear
hot plasma and gyrokinetic theories agree when the gyrokinetic
approximation is valid, we will continue to use the solution
of the gyrokinetic model to specify the path of the cascade in the
$(k_\perp,k_\parallel)$ plane. This is probably at least qualitatively 
correct because cyclotron damping does not in fact become large.

To quantify the conditions where the cyclotron resonance terms becomes
important we define a threshold value of $\kperp$ to be the value
along the cascade path at which the solutions of the gyrokinetic and
hot plasma dispersion relations diverge by 10\%.  This threshold is
plotted (solid line) in
\figref{fig:cyc_thresh_num} for a range of $\beta_i$ with parameters
$T_i/T_e = 1$ and $k_0 \rho_i = 10^{-4}$. We probe the sensitivity of
the threshold $\kperp$ to our assumptions by plotting the same 10\%
threshold for the path in the $(k_\perp,k_\parallel)$ plane calculated
from the gyrokinetic model with {\em no} damping (dashed line).  The
dotted line is an estimate of this threshold determined by solving
the $k_\perp \rho_i \gg 1$ limit of
\eqref{eq:womegai2} with $\omega/\Omega_i = 1$ and $\epsilon =
\epsilon_0$ to find
\begin{equation}
k_\perp \rho_i \simeq (k_0 \rho_i)^{-1/4} \beta_i^{5/8} \left[
1+\frac{2}{\beta_i (1+ T_e/T_i)}\right]^{1/4}
\label{eq:rough}
\end{equation}
(this rough estimate is only plotted over the typical range of
$\beta_i$ in the solar wind near Earth).

The solid line in \figref{fig:cyc_thresh_num} represents our best
estimate of the threshold at which the cyclotron resonance becomes
important and the gyrokinetic approximation fails. If the effective
damping of the fluctuations in the nonlinear turbulence is less than
the linear damping rates, the true threshold will fall within the
vertically shaded region. This region is bounded from below by the
short-dashed line---our estimate of the threshold with no
damping---and represents the most conservative estimate of the limit
of applicability of gyrokinetics. Thus, throughout the diagonally
shaded region, gyrokinetics is certainly a valid description of the
turbulent dynamics in the plasma, and this region covers a substantial
part of the parameter range of interest in the study of turbulence in
the solar wind.

The cyclotron resonance has a strong effect on the cascade only when
it is reached at a relatively low wavenumber, $k_\perp \rho_i \lesssim
1$; otherwise, the cascade is likely to be rather heavily damped
before reaching the cyclotron frequency, leaving little energy
available for perpendicular heating.  Equation~(\ref{eq:rough}) provides
a rough estimate of how the wavenumber threshold in the undamped
case scales with the plasma and system parameters.  This \emph{undamped} threshold
has little dependence on the temperature ratio $T_i/T_e$.  Thus, the
cyclotron resonance occurs early on in the dissipation range only in plasmas with
$\beta_i \lesssim 0.01$ or when the isotropic driving wavenumber is
much larger than our fiducial estimate $k_0 \rho_i \gg 10^{-4}$. 

\subsection{Effect of the Ion Cyclotron Resonance}
\label{sec:conversion}

Let us now discuss what happens if the cyclotron resonance is reached
in the dissipation range ($\kperp\rho_i>1$) and before the kinetic
\Alfven wave cascade is converted into heat by Landau damping
($\kperp\rho_e<1$).  The idealized asymptotic version of the
appropriate linear theory (the case $\kperp\rho_i\gg1$,
$\kperp\rho_e\ll1$ at the cyclotron resonance) can be worked out
analytically from the general hot plasma dispersion relation---this is
done in Appendix~\ref{app:cycdamping}.  The key points that this
calculation demonstrates are
\begin{itemize}
\item the cyclotron effects only matter when the 
frequency of the fluctuations is very close to one of the resonances: 
for integer $n\ge1$, 
$(\omega-n\Omega_i)/(|\kpar| v_A\kperp\rho_i)\sim {1/\sqrt{\kperp\rho_i}}$; 
\item away from the resonances, the ions effectively have a Boltzmann response 
and kinetic \Alfven waves (weakly damped by the electron Landau
resonance) exist both below the ion cyclotron frequency
($\omega<\Omega_i$) and between the resonances
($n\Omega_i<\omega<(n+1)\Omega_i$);
\item at the resonances, there is a conversion between 
kinetic \Alfven waves and ion Bernstein waves---electrostatic waves
whose frequencies are close to $n\Omega_i$
\citep[cf.][]{LiHabbal:2001};
\item because $\kperp\rho_i\gg1$, the ion cyclotron damping is 
exponentially weak ($\sim e^{-\kperp\rho_i}$) and very localized in frequency.
\end{itemize}
These points are illustrated in \figref{fig:app}.  Together, they
imply that the effect of passing through the cyclotron resonance on
the kinetic \Alfven wave cascade is unlikely to be drastic: in order
to be subjected to even a weak cyclotron damping, the fluctuations
have to couple nonlinearly into a rather narrow frequency band around
$\Omega_i$; it appears more likely that most of the turbulent energy
would leapfrog the resonance altogether and continue as a kinetic
\Alfven wave cascade, until converted into heat by electron Landau
damping (at $\kperp\rho_e\sim1$). However, it is probably prudent to
acknowledge that linear calculations such as the one presented in
Appendix~\ref{app:cycdamping} may have limited utility in
predicting nonlinear behavior.

It is important to understand that the calculation in
Appendix~\ref{app:cycdamping} considers an idealized asymptotic
solution. It is known that the precise mode conversion and damping
properties at the cyclotron resonances depend rather sensitively on
the plasma parameters: such behavior at oblique ($k_\perp >
k_\parallel$) propagation was explored in detail by
\citet{LiHabbal:2001}.  To demonstrate some of the non-asymptotic
possibilities that may be present in the solar wind, it is perhaps
useful to consider here a rather extreme low-$\beta_i$ example in
which the ion cyclotron resonance is reached quite close to the
beginning of the dissipation range.

In \figref{fig:gk_hp_w_kpar}, we present the cascade path on the
$(k_\perp,k_\parallel)$ plane (solid) and the frequencies along that
path calculated by numerically solving 
both the linear gyrokinetic (dashed) and hot plasma (dotted)
dispersion relations for plasma parameters $\beta_i=0.03$, $T_i/T_e=3$, and
$k_0 \rho_i= 2 \times 10^{-4}$ (chosen to highlight mode conversion features). 
Two branches of the hot plasma
dispersion relation are plotted: in the MHD limit $k_\perp \rho_i \ll
1$, these correspond to the MHD \Alfven wave and MHD fast wave. As
discussed in \secref{sec:turb}, the MHD \Alfven wave transitions to a
kinetic \Alfven wave at $k_\perp \rho_i > 1$. The corresponding solution of the
gyrokinetic dispersion relation, which does not include the cyclotron
resonance, continues right through $\omega=\Omega_i$. 
The effect of the cyclotron resonance on the \Alfven branch
of the hot plasma solution is that, as $\omega \rightarrow \Omega_i$,
the kinetic \Alfven wave converts into an ion Bernstein wave at $k_\perp
\rho_i \simeq 3$. The fast wave has a similarly interesting behavior
in this example. At $k_\perp \rho_i \simeq 0.3$, the fast wave
converts to an ion Bernstein wave; eventually this same branch,
at $k_\perp \rho_i \simeq 5$ connects onto a continuation of the
kinetic \Alfven wave in the range $\Omega_i < \omega < 2\Omega_i$
before becoming an $n=2$ ion Bernstein
wave at $k_\perp \rho_i \simeq 8$. 

In a turbulent cascade, different wave modes can couple nonlinearly,
but if the nonlinear transfer is local in wavenumber, the coupling
will be strong only when those modes have similar wavevectors and
frequencies. Thus, in our example, at $k_\perp \rho_i \simeq 0.1$, the
two branches (the fast and \Alfven modes) have frequencies that differ
by an order of magnitude, so the coupling is likely to be weak; but,
at $k_\perp\rho_i \simeq 4$, their frequencies differ by only a factor
of 2, so the coupling may be much more efficient. In this case, the
coupling may enable the turbulent energy of the kinetic \Alfven wave
on the low-frequency branch to cascade nonlinearly to the continuation
of the kinetic
\Alfven wave on the high-frequency branch. The efficiency of such a
process and the amount of perpendicular cyclotron heating that can
occur as the cascade passes through the resonance is likely to depend
on the damping rates of the modes involved.  In
\figref{fig:gk_hp_gwie2} we plot the damping rates onto ions and
electrons for the solutions of the gyrokinetic (dashed) and hot plasma
(dotted) dispersion relations for the same parameters as in
\figref{fig:gk_hp_w_kpar}.  In the gyrokinetic case, the ion Landau
damping falls below the plotted range, while the electron Landau
damping is significant (indeed, it is this damping that served as the
primary kinetic input in our cascade model).  For the low-frequency
(\Alfven) branch (upper panel), there is a rise of ion cyclotron
damping at $k_\perp\rho_i \simeq 3$ and a slight decrease in the
electron Landau damping.  For the high-frequency (fast) branch (lower
panel), there is a narrow peak of ion cyclotron damping at
$k_\perp\rho_i \simeq 0.16$, the point where the fast wave frequency
crosses $\omega=\Omega_i$. Two more ion cyclotron damping peaks occur
in wavenumber ranges where the fast branch takes on the
characteristics an ion Bernstein wave.  (Similar cyclotron damping
peaks for the low- and high-frequency branches are seen in the more
asymptotic case considered in Appendix~\ref{app:cycdamping}; see
\figref{fig:app}). Thus, if the kinetic
\Alfven wave cascade reaches the ion cyclotron frequency, it is
possible for some of the energy to go into perpendicular ion heating
via coupling to the ion Bernstein wave modes. However, for parameters
typical of the slow solar wind, and even for the rather extreme
low-$\beta_i$ example considered here, the ion cyclotron damping is
weak because it only sets in at $k_\perp \rho_i \gtrsim 1$; in
\figref{fig:gk_hp_gwie2} it remains at least an order of magnitude
below the electron Landau damping.

As noted above, when the cascade reaches the ion cyclotron frequency,
the path of the cascade on the $(k_\perp,k_\parallel)$ plane,
calculated using the gyrokinetic cascade model, is unlikely to be
quantitatively correct. Nonetheless, we can tentatively examine the
differences in the resulting magnetic energy spectra between the
gyrokinetic and hot plasma cascade models (defined in
\secref{sec:cyclo}).  These are plotted in \figref{fig:diss_gk_hp} for
the same parameters as the example in Figures~\ref{fig:gk_hp_w_kpar}
and \ref{fig:gk_hp_gwie2}. Because the ion cyclotron resonance does not
significantly alter the total damping rates (see the upper panel of
\figref{fig:gk_hp_gwie2}), the resulting spectra differ very little,
both producing an effective dissipation range spectral index near
$-4$. This suggests that the effect of the ion cyclotron resonance
cannot be identified based solely on the magnetic fluctuation energy
spectrum; an examination of the underlying dynamics in concert with a
forward modeling approach, like the wave phase velocity study in
\secref{sec:phasevel}, is necessary if one hopes to distinguish
between ion cyclotron and electron Landau damping and, consequently,
between ion and electron heating.

\section{Discussion}
\label{sec:discuss}

The gyrokinetic cascade model developed in this paper is meant to
describe in a simple way the physical mechanism---the nonlinear
cascade through the MHD \Alfven wave and kinetic \Alfven wave
regimes---by which the vast reservoir of energy contained in
large-scale turbulent eddies in the solar wind cascades to high
wavenumber and frequency and ultimately dissipates due to
collisionless damping. Our aim is to reconcile observational data with
theoretical ideas about the energy transfer to the small dissipative
scales, and eventually to thermal energy of the plasma particles. In
that light, this section contains a lengthy discussion of the
limitations of our model, followed by a discussion of its relation to
previous work.
 
\subsection{Limitations of Our Model}
\label{sec:uncert}

One of the key assumptions of our model is that the turbulence can be
treated as if it were driven isotropically at some large scale,
characterized by an effective isotropic driving wavenumber $k_0
\rho_i$. Two exceptions can be made to this assumption: 
the turbulence may not be driven isotropically, and it may be driven
at more than one scale. 

\paragraph{Possibility of Anisotropic Driving} 
To dispense with this question, we recall that, as we
noted in \secref{sec:turb}, our model does not strictly
require that the turbulence is driven isotropically. The critical
balance path in the $(k_\perp,k_\parallel)$ plane, as plotted in
\figref{fig:kplane}, simply requires a single point to fix the
relation---we use the parameter $k_0 \rho_i$ as a convenient
characterization of that point. The anisotropy of the slow solar wind
turbulence at $k_\perp \rho_i \sim 10^{-3}$, measured by
\citet{Dasso:2005}, supplies the necessary data point to pin down the
scale-dependent anisotropy of the GS theory; we determine the
characteristic value of of $k_0 \rho_i$ based on that point. A strong
test of this model would be a measurement of the anisotropy, similar
to that in \citet{Matthaeus:1990}, at a smaller scale around $k_\perp
\rho_i \sim 0.1$.  At this wavenumber the anisotropy is predicted to
be much stronger with $k_\perp/k_\parallel \sim 10$.

A note is, perhaps, necessary regarding our use of frequencies and
damping rates from the gyrokinetic dispersion relation for the entire
turbulent cascade even though gyrokinetics is obviously not valid at
the isotropic driving scale. While our model undoubtedly does not
describe the driving scale well, the effect of driving is probably
well modeled by the simple idea of a constant energy input: the
damping rate is negligible at the driving scale and the \Alfven
frequency gives a good order-of-magnitude approximation to the
nonlinear cascade rate, so the values from the gyrokinetic theory are
satisfactory.  As we noted in \secref{sec:appgk}, the assumptions
behind the gyrokinetic theory are ever better satisfied as we proceed
deeper into the inertial range.

\paragraph{Possibility of Small-Scale Driving}
The second exception is that energy may be injected at multiple
scales. In \secref{sec:drivescale}, we argued that our estimated
effective driving scale was consistent with the turbulence driven at
the Sun and then decaying in transit to 1 AU.  It is unlikely that
there are large-scale energy sources affecting the turbulence during
this transit.  It has, however, been suggested that Kelvin-Helmholtz
instabilities arising from shear between fast and slow solar wind
streams could continually drive turbulence in the solar wind
\citep{Coleman:1968,Roberts:1987b,Roberts:1991,Roberts:1992,Zank:1996}; 
such a mechanism could inject energy at scales smaller than 
our estimated isotropic driving scale. 
However, magnetic fluctuation frequency spectra for the slow
solar wind typically show a very clean $-5/3$ spectrum all the way to
the low frequencies associated with our fiducial driving wavenumber
$k_0 \rho_i \simeq 10^{-4}$
\citep{Tu:1990b,Grappin:1990,Tu:1995}, suggesting a continuous
inertial range below the energy injection scale.  If significant
energy were injected at smaller scales, one would expect to see
structure in the magnetic energy spectrum at the associated
frequencies. The lack of such structure argues against significant
energy injection at smaller scales. Hence, we conclude that a single
effective isotropic driving wavenumber $k_0 \rho_i$ is a reasonable
assumption.

\paragraph{Plasma Instabilities} 
Another potential source of small-scale energy injection neglected in
our model is the possible presence of temperature anisotropy
instabilities, e.g., the mirror or firehose instabilities.  Adiabatic
expansion of the solar wind drives ions towards $T_{i \perp} < T_{i
\parallel}$, which should trigger firehose instabilities
\citep{Scarf:1967}; indeed, there exists observational evidence that
the solar wind ion temperature anisotropies are constrained by the
firehose marginal stability criterion \citep{Eviatar:1970,Kasper:2002}.  These
instabilities may inject energy into the cascade at
$k_\parallel\rho_i\sim 1$ and $\omega \sim \Omega_i$.  It is not
possible yet to determine whether such energy injection is present in
the data.  The role of kinetic instabilities will be investigated in
future studies.

Besides the presence of a single isotropic driving scale, 
two other key assumptions were made in the construction of our model: 

\paragraph{Critical Balance}
We followed \citet{Goldreich:1995} in 
assuming that the turbulent cascade maintains a state of
critical balance, $\omega=\omega_{nl}$. Imposing critical balance and
choosing a driving wavenumber $k_0 \rho_i$ enables one to determine
the frequency of the cascade with respect to the ion cyclotron
frequency. Critical balance is used to approximate the
energy flow on the $(k_\perp,k_\parallel)$ plane as a
one-dimensional path, as shown by the solid line in
\figref{fig:kplane}; the turbulent energy is, however, expected not to
be restricted to this line but to fill the shaded region for which
$\omega \lesssim \omega_{nl}$ (the hydrolike modes in the nomenclature
of \citet{Oughton:2004}); simulations of MHD turbulence support this
notion
\citep{Cho:2000,Maron:2001,Cho:2002,Cho:2003,Oughton:2004}.  An
extension of this cascade model to two dimensions, allowing energy to
fill modes throughout the $(k_\perp,k_\parallel)$ plane, is required
to relax this assumption; we do not believe such a treatment would
substantially alter the conclusions of this paper. Ultimately, the 
premise of critical balance will be evaluated by 
the nonlinear gyrokinetic simulations of magnetized turbulence. 

\paragraph{Linear Damping Rates for Nonlinear Cascade} 
We assumed that the damping rates for linear eigenmodes are
appropriate for the nonlinear turbulent cascade in which the energy in
a wave mode is transferred to higher wavenumbers within a single wave
period. \citet{Li:2001} call into question the validity of this linear
damping approach, citing results in contrast to observation; for a
more detailed comparison of this paper to their work, see
\secref{sec:previous}. We find that our results are consistent with a
range of observations, but we must leave the final judgment to
nonlinear gyrokinetic simulations.

Finally, we ignored a number of potentially significant physical
effects.  One of them was the possible presence of small-scale plasma
instabilities briefly mentioned above. Some of the others are
discussed below.

\paragraph{Radial Expansion of the Solar Wind}
Not only does this have the effect of
altering the parameters of the plasma---for example, the plasma
transits from the low $\beta_i$ corona to the order unity $\beta_i$
solar wind at 1~AU---but the expansion may also affect the turbulent
fluctuations. We focus here on the properties of the slow solar wind
at 1~AU, but admit a possible influence of this expansion on our
results. There exists a substantial literature focusing on the spatial
transport and evolution of the solar wind throughout the heliosphere
\citep[see][for a review]{Tu:1995}. The turbulence in these models is
usually specified locally by prescription.
Our model is also local, so the effect  
of the expansion on the turbulence is not addressed. 

\paragraph{Imbalanced Turbulence}
Early measurements of the fast solar wind at large scale suggested it
was dominated by anti-sunward propagating \Alfven waves
\citep{Belcher:1971}. Subsequent measurements in the fast solar
wind found that anti-sunward Els\"asser energy fluxes can exceed
sunward fluxes by more than an order of magnitude; in the slow wind,
the imbalance is reduced, sometimes showing equal energy in both
directions \citep{Tu:1990a,Tu:1990b,Grappin:1990,Tu:1995}. 
Our model only had one flux quantity $\epsilon$, i.e., 
it assumed a balance of Els\"asser energy. An imbalanced turbulence 
requires a somewhat more complicated treatment
\citep{Dobrowolny:1980,Grappin:1983,Politano:1989,Hossain:1995,Lithwick:2003,Lithwick:2007}; 
thus, a more general cascade model would need to include an additional
parameter to account for the imbalance between the fluxes.  While this
means that our model is likely not appropriate for the fast solar
wind, we believe that it is applicable to the slow solar wind when the
Els\"asser energy fluxes are balanced.

Note also that our model does not address the implications of the
conservation of magnetic and cross helicities in MHD.

\paragraph{Compressive Modes}
Compressive fluctuations, such the MHD slow, entropy, and fast modes, 
or, more generally, the density and field strength fluctuations, 
are left out of our model. Observational evidence suggests
less than 10\% of the power resides in compressive modes
\citep{Belcher:1971}. Linear theory predicts that these modes are
damped when $k_\parallel\lambda_{{\rm mfp}i} \gtrsim 1$ in the weakly collisional solar wind
\citep{Braginskii:1965,Barnes:1966}. However, it has been suggested that
a fast wave cascade may be able to transfer energy to high frequencies
where it can heat ions perpendicular to the magnetic field
\citep{Chandran:2006}. The cascade of slow and entropy modes
is also under investigation \citep{Schekochihin:2007}---it is not clear that the 
parallel cascade of these modes is similar to the \Alfven wave cascade. 
If a cascade of compressive fluctuations 
carries a significant amount of energy to the ion Larmor radius 
scale, this can contribute to the kinetic \Alfven wave cascade 
in the dissipation range, thus altering somewhat the energetics 
of our model. 

\paragraph{Kinetic Entropy Cascade}
We have referred to collisionless kinetic damping of electromagnetic
fluctuations as the mechanism that leads to plasma heating. The way
this heating occurs is, in fact, quite complex. Collisionless damping
is an energy conversion mechanism that transfers the electromagnetic
energy of the MHD and kinetic \Alfven wave cascades into purely
kinetic cascades of ion and electron entropy fluctuations---cascades
that occur in phase space and that take the energy to sufficiently
small spatial and velocity scales for the collisions to render the
``collisionless'' damping irreversible and to effect the heating. How
these cascades work is explained in \citet{Schekochihin:2007}. An
important point here is that these cascades should have observational
signatures, in particular, steep power-law spectra of low-frequency
electric and magnetic fluctuations \citep{Schekochihin:2007}.  A
superposition of these kinetically induced fluctuations with the
kinetic \Alfven wave cascade may contribute to the wide spread in the
observed spectral index in the dissipation range.
\subsection{Relation to Previous Work}
\label{sec:previous}
The cascade model proposed here is similar to that constructed by
\citet{Zhou:1990c} and later used by \citet{Li:2001} in a study of the
dissipation of solar wind turbulence. The \citet{Zhou:1990c} model
employed a diffusion operator for the nonlinear cascade of energy in
wavenumber space \citep{Leith:1967} to develop an {\em isotropic}
model for turbulence in the MHD regime. This model was used by
\citet{Li:2001} to determine the characteristics of the resulting
energy spectrum in the dissipation range when collisionless damping by
the cyclotron and Landau resonances was considered; they concluded
that these mechanisms could only cause a steep cutoff in wavenumber
rather than the power-law seen in observations. We come to a different
conclusion for reasons that we shall now explain.

First, the \citet{Li:2001} model did not assume that the cascade
frequency was determined by critical balance, so the change of
dynamics leading to a faster cascade beginning at $k_\perp \rho_i \sim
1$ was neglected. Second, they followed the cascade along constant
angles in $(k_\perp,k_\parallel)$ plane of $\theta=0^\circ$,
$30^\circ$, $45^\circ$, and $60^\circ$, effectively assuming either
isotropy or a scale-independent anisotropy. On the other hand, our
critically balanced model predicts that at $k_\perp \rho_i \sim
1$---where collisionless damping first becomes significant---the
anisotropy should be $k_\parallel/k_\perp \sim 0.1$, leading to an
angle $\theta
\gtrsim 85^\circ$.  For these nearly perpendicular modes, the
frequency remains low, $\omega\ll
\Omega_i$, so the \Alfven wave cascade is damped only by the Landau
resonance, and the damping is less vigorous.  Hence, we find a cutoff
due to collisionless damping that is much less sharp than that found
by \citet{Li:2001}, and that is consistent with \emph{in situ}
measurements. 

In a related work, \citet{Gary:2004} also found that ion cyclotron
damping produced a steep damping inconsistent with
observations. Furthermore, they noted that, in the limit that
$k_\parallel < k_\perp$, electron Landau damping would dominate
collisionless damping with a more gradual onset.  They suggested a
qualitative picture that, in this case, the location of the spectral
break between the inertial and dissipation ranges would be dependent
on the turbulent energy cascade rate. We agree with their prediction
of the more gradual onset of Landau damping, but believe the spectral
break is most often caused by a change in dynamics at $k_\perp \rho_i
\sim 1$, a feature not included in their paper.

\citet{Smith:2001b} investigated an unusual period of solar wind 
turbulence with $\beta_i \ll 1$, a condition in which the ion inertial
length and the ion Larmor radius are well separated; they concluded
that the dissipation processes occur at the ion inertial length and
that this finding is inconsistent with dissipation due to the ion
cyclotron resonance.  Here we compare our hot plasma cascade model
predictions directly to interval (c) of \citet{Smith:2001b}. From
Figure~1 of their paper, we estimate the following mean values of the
parameters over interval (c): $B_0 \simeq 6$~nT, $V_{\rm sw}
\simeq 350$~km/s, $T_i \simeq 6 \times 10^4$~K, and $\beta_i \simeq 0.01$. 
We employ our fiducial driving wavenumber $k_0\rho_i = 10^{-4}$, and
take the mean electron temperature for the slow solar wind as $T_e
\simeq 1.3 \times 10^5$ K \citep{Newbury:1998}, giving a temperature
ratio $T_i/T_e = 0.5$.  For these parameters, the ion Larmor radius is
$\rho_i \simeq 5.5 \times 10^6$~cm. As presented in the upper panel of
\figref{fig:smith}, the steady-state magnetic energy spectrum from the
hot plasma cascade model breaks from the $-5/3$ spectral index at a
value of $k_\perp \rho_i \simeq 0.25$; we transform from the
wavenumber spectrum generated by our model to a frequency spectrum in
the satellite frame by the Taylor hypothesis, using $f= (k_\perp
\rho_i) v_{sw}/(2 \pi \rho_i)$ to find a breakpoint frequency of
$f\simeq 0.25 $ Hz.  \citet{Smith:2001b} measure the breakpoint
frequency to be $f \simeq 0.168$ Hz; the hot plasma cascade model
therefore predicts a breakpoint frequency that is roughly consistent
with the observations. In the lower panel of \figref{fig:smith}, it is
shown that electron Landau damping is responsible for the break. In
this case, the damping is sufficient to prevent the cascade from
reaching the ion cyclotron frequency; thus, ion cyclotron damping is
negligible, falling below the plotted range of the lower panel. We
note here that the gyrokinetic cascade model gives results nearly
indistinguishable from those presented in \figref{fig:smith} for the
hot plasma cascade model; this is not surprising since the cyclotron
resonance plays a negligible role.  It is worth noting that the
wavenumber associated with the breakpoint, when cast in terms of the
ion inertial length, is $k_\perp d_i \simeq 2.5$, so we agree with the
\citet{Smith:2001b} conclusion that the breakpoint occurs at the ion
inertial length scale.  The salient point here is that, although the
magnetic energy spectrum breakpoint often occurs at $k_\perp \rho_i
\sim 1$ due to the change in linear physics from the MHD to the
kinetic \Alfven wave regime, strong damping may shift the breakpoint
to a lower value of $k_\perp \rho_i$.

\subsection{Application to the Solar Corona}
\label{sec:corona}
Although this paper has focused on anisotropic turbulence in the solar
wind, such turbulence is also expected to be present in the solar
corona.  One of the key problems for the solar corona is the
identification of the mechanism by which the coronal plasma is heated
to yield the observed minority ion temperatures higher than proton
temperatures $T_{\mbox{ion}}>T_p$ and large ion and proton temperature
anisotropies $T_\perp \gg T_\parallel$
\citep{Kohl:1997,Kohl:1998,Antonucci:2000,Kohl:2006}. In the spirit of 
a long line of previous investigations
\citep{Hollweg:1986,Isenberg:1990,Li:1999,
Matthaeus:1999,Dmitruk:2001,Dmitruk:2002,Cranmer:2003,Cranmer:2005,Cranmer:2007},
we brifely address what our cascade model predicts for ion heating in
the solar corona.

The first caveat is that the cascade model described in this paper
assumes a balance of the oppositely-directed Els\"asser energy fluxes
along the magnetic field line. Therefore, the predictions of this
simple model cannot be applied to the coronal hole regions in which
the field lines are open and the anti-sunward \Alfven wave energy flux
dominates; the conclusions here will only apply to regions of closed
field lines in which the turbulence demonstrates balanced energy
fluxes. For the solar corona at a heliocentric radius $r=2 R_\odot$,
we take the following plasma parameters: magnetic field magnitude
$B_0=0.8$~G, number density $n=3.7 \times 10^5$~cm$^{-3}$, electron
temperature $T_e=8 \times 10^5$~K, and proton temperature $T_e=2
\times 10^6$~K. These plasma conditions give an ion Larmor radius
$\rho_i=2.4 \times 10^3$~cm, an ion cyclotron frequency $\Omega_i=7.7
\times 10^3$~rad/s, an ion plasma beta $\beta_i=0.004$, and a
temperature ratio $T_i/T_e=2.5$.

In addition to these parameters, for the hot plasma cascade model we
must also specify the driving by supplying the characteristic
isotropic driving wavenumber $k_0 \rho_i$. From the model of
\citet{Cranmer:2005}, we find that, at a radius of $r=2 R_\odot$,  
the driving energy due to photospheric motions peaks at a period of
$P=120$~s. Using the dispersion relation for \Alfven waves
$\omega=k_\parallel v_A$, this gives a parallel driving length scale
$L_\parallel=3.4\times 10^{10}$~cm. At this radius, the
\citet{Cranmer:2005} model estimates the perpendicular driving 
scale $L_\perp = 1.3\times 10^9$~cm (using their best value of
$\Lambda=0.35$ in their equation~[51]); for the period $P=120$~s, this
implies a perpendicular velocity of $v_\perp = 110$~km/s.  The
effective isotropic driving wavenumber can be found using
\eqref{eq:k0}, giving the result $k_0 \rho_i= 7 \times 10^{-10}$. Note 
that, although the length scale associated with $k_0$ is $L_0=2 \times
10^{13}$~cm, there are no motions at this scale; the driving occurs
anisotropically at $L_\parallel > L_\perp$ and the value of $k_0
\rho_i$ calculated here is simply a convenient way to specify the critical balance 
relation determined by the driving mechanism, as explained in
\secref{sec:turb}.

The heating resulting from the damping of the turbulence can occur via
either the Landau or cyclotron resonances in the hot plasma
cascade model; Landau damping leads to an increase of the parallel
temperature, cyclotron damping to an increase of the perpendicular
temperature. For the low $\beta_i$ conditions in the solar corona,
Landau damping onto the ions is negligible, so all heating is due to
either electron Landau damping or ion cyclotron damping. Using
\eqref{eq:womegai2} to estimate the frequency of the cascade at $k_\perp
\rho_i =1$ (and neglecting damping by taking $\epsilon=\epsilon_0$), we find
$\omega/\Omega_i = 1.7 \times 10^{-2}$; this suggests ion cyclotron
damping will be negligible in this case. The steady-state solution of
the hot plasma cascade model for these parameters confirms that the 
\Alfven wave cascade driven by photospheric motions results in only
parallel electron heating via the Landau resonance. In fact, for all
values of $k_0 \rho_i < 10^{-5}$, the turbulent cascade fails to reach
the ion cyclotron frequency and therefore no ion heating occurs.
Significant perpendicular ion heating via the balanced, \Alfven wave
cascade requires an isotropic driving scale $L_0\le 1.5 \times
10^7$~cm, a condition not satisfied in the solar corona.  In
conclusion, this turbulent cascade model cannot account for the
perpendicular ion heating observed in the solar corona.

\section{Conclusions}
\label{sec:conc}
In the study of solar wind turbulence, a central issue is to
understand the mechanisms by which the energy contained in large-scale
turbulent motions is transferred to smaller scales and higher
frequencies where it is eventually deposited as thermal energy in the
plasma. In \secref{sec:model} we develop a simple model that traces
the anisotropic cascade of \Alfven wave energy from the MHD to the
kinetic \Alfven wave regime. The model is based on the assumption of
locality of energy transfer in wavenumber space, the conjecture of a
critical balance between the nonlinear interaction and linear
propagation timescales, and the premise that the linear kinetic
damping rates determine heating rates even in the presence of the
nonlinear cascade.  It provides a guide for the interpretation of
observational data in the balanced, slow solar wind and will be useful in
connecting these observations to nonlinear gyrokinetic simulations of
turbulence.

The anisotropic cascade model makes several predictions about the
turbulent magnetic energy spectrum measured in the balanced, slow solar
wind---these are presented in \secref{sec:spec1}--\secref{sec:kc}.
The cascade is weakly damped in the \Alfven wave regime $k_\perp
\rho_i \ll 1$ and the spectrum has the familiar $k^{-5/3}$ slope.  In
the dissipation range, at wavenumbers above the observed break, the
magnetic energy spectrum undergoes a slow exponential cut-off. We show
that the measured power-law behavior of the spectrum in the
dissipation range may be an artifact of limited magnetometer
sensitivity (except in the case of very weak damping) and that over
the typical range of parameters $\beta_i$ and $T_i/T_e$ in the solar
wind, the varying strength of Landau damping naturally reproduces the
observed variation of the effective spectral index in the dissipation
range from $-7/3$ to $-4$ (\secref{sec:noise}-\secref{sec:index}).
The spectral break between the inertial range and dissipation range
power laws occurs at the ion Larmor radius $k_\perp \rho_i \simeq 1$
in plasmas with weak damping due to the dispersive nature of the
kinetic \Alfven wave; in the presence of strong damping, it may occur
at $k_\perp \rho_i \lesssim  1$ due to damping by the Landau
resonance and/or the ion cyclotron resonance. See \figref{fig:gw} for
a plot of the damping rate as a function of plasma parameters at
$k_\perp \rho_i =1$.

In \secref{sec:appgk} we examine the applicability of gyrokinetics to
turbulence in the balanced, slow solar wind.  At scales somewhat below the
driving scale, the anisotropy of the turbulence, $k_\parallel <
k_\perp$, is supported by observations \citep{Dasso:2005} in the slow
solar wind.  Given the observed anisotropy, we find that a gyrokinetic
description becomes valid from a scale a couple of orders of magnitude
below the driving scale up to the point where the fluctuation
frequency approaches the cyclotron frequency, $\omega \simeq
\Omega_i$. The driving scale of the turbulence in the slow solar wind
is estimated to be $L\simeq 2 \times 10^{11}$~cm
(\secref{sec:drivescale}), in agreement with three other independent
determinations.  For typical solar wind parameters, we find that
$k_\parallel \ll k_\perp$ and $\omega \ll \Omega_i$ at a perpendicular
wavenumber $k_\perp \rho_i \sim 1$ (\secref{sec:aniso}); therefore,
this transition regime is well described by gyrokinetics. Wave phase
velocity evidence from \emph{in situ} observations \citep{Bale:2005}
argues for the existence of a kinetic
\Alfven wave cascade in the dissipation range (which is captured in
our model); the measurements are inconsistent with the onset of ion
cyclotron damping (\secref{sec:phasevel}).

The gyrokinetic cascade and the resultant particle heating through
Landau damping, however, cannot be the whole story.  The selective
perpendicular heating of minor ions and the breaking of adiabatic
invariance for ions implied by observational data, particularly in the
fast wind, indicate that fluctuations at or above the cyclotron
frequency are present in the solar corona and solar wind.  Although a
mechanism not incorporated into our model may be responsible---for
example, heating by fast waves
\citep{Chandran:2006} or velocity-space instabilities that transfer
energy from the parallel to the perpendicular direction
\citep{Scarf:1967,Eviatar:1970,Kasper:2002,Marsch:2006}---we may employ our model 
to determine under what conditions the gyrokinetic cascade reaches the
ion cyclotron frequency. For plasma parameters typical of the solar
wind and the driving scale estimate $L\simeq 2 \times 10^{11}$~cm, the
gyrokinetic cascade model generally reaches the ion cyclotron
frequency at $k_\perp \rho_i \gg 1$, by which point  Landau damping 
has removed much of the cascade energy.  In addition, damping of
the cascade can slow the increase of the frequency with perpendicular
wavenumber, possibly preventing the cyclotron frequency from being
reached at all, as shown in \figref{fig:undamped}. 
A strong constraint on the wavenumber
threshold at which the cyclotron resonance becomes important---the
same threshold at which the gyrokinetic approximation is
violated---is determined by a comparison of the gyrokinetic and
hot plasma cascade models. The regime of validity for gyrokinetics is
given by the shaded region in
\figref{fig:cyc_thresh_num} and covers most of the
parameter range of interest in the study of turbulence in the solar
wind. In the range of parameters relevant to the solar wind, this
threshold roughly scales as $k_\perp \rho_i \sim (k_0\rho_i)^{-1/4} \beta_i^{5/8}$
(\secref{sec:cyclo}). Therefore, the cyclotron resonance can 
have a significant effect on the dissipation-range dynamics 
only in plasmas with $\beta_i < 0.01$ or
in systems driven isotropically at scales $L \ll 2\times10^{11}$~cm. 
If the cyclotron resonance is reached within the dissipation 
range before the electron Landau damping becomes dominant, 
there is a conversion between kinetic \Alfven waves and ion 
Bernstein waves in the narrow frequency band around the 
resonance. However, the amount of cyclotron damping associated 
with this conversion is quite small and the narrowness
of the resonance suggests that the kinetic \Alfven wave cascade 
may in fact continue virtually unimpeded even at $\omega>\Omega_i$
(see \secref{sec:conversion} and Appendix~\ref{app:cycdamping}). 

In \secref{sec:corona}, we apply this cascade model to the problem of
ion heating in the solar corona. For the low $\beta_i$ conditions of
the corona, the dissipation of turbulence can cause parallel heating
of the electrons via the Landau resonance or perpendicular heating of
the ions via the ion cyclotron resonance.  We conclude,
however, that the highly anisotropic driving due to photospheric
motions results in only parallel electron heating via the Landau
resonance; only for an effective isotropic driving scale $L_0\le 1.5
\times 10^7$~cm, a condition not satisfied in the corona,  will the 
\Alfven wave cascade as modeled here lead to substantial perpendicular 
ion heating.

Although our simple cascade model allows us to produce plausible
predictions for magnetic-energy spectra, reaching a definitive
understanding of turbulence in the solar wind will require nonlinear
simulations of magnetized plasma turbulence in the weakly collisional
regime.  A program of gyrokinetic simulations is currently underway
\citep{Howes:2007b,Howes:2007e}.  These simulations will both test the
validity of the ideas that underpin the approach taken in this paper
and probe the physics beyond the scope of our simple model, in
particular the complicated dynamics at the scale of the ion Larmor
radius $k_\perp
\rho_i \sim 1$. To provide detailed information for comparison to
\emph{in situ} measurements, gyrokinetic simulations will calculate
particle heating rates, magnetic and cross helicities, and nonlinear
energy transfer rates in imbalanced cascades. The cascade model
proposed in this paper is a first step in connecting the results of
nonlinear gyrokinetic simulations to the observations in the solar
wind and in assessing the applicability of gyrokinetics to this
problem.


\begin{acknowledgments}
We thank S.~Bale for useful discussions.  G.G.H. was supported by the
DOE Center for Multiscale Plasma Dynamics, Fusion Science Center
Cooperative Agreement ER54785. G.G.H, S.C.C., and W.D. thank the Aspen
Center for Physics for their hospitality. E.Q. and G.~G.~H. were
supported in part by the David and Lucille Packard Foundation.
A.~A.~S. was supported by an STFC Advanced Fellowship and an RCUK 
Academic Fellowship.
\end{acknowledgments}

\begin{appendix}
\section{Plasma Dispersion  Relation,  
Mode Conversion and Damping 
at the Ion Cyclotron Resonance}
\label{app:cycdamping}

The general eigenvalue problem for linear perturbations of a thermal 
plasma is 
\beq
{\bf\hat D}\cdot\delta{\bf E} = 0.
\eeq
where,  
if we choose the coordinate frame so that $z$ is the 
direction of the mean magnetic field and $x$ is the direction of 
${\bf k}_\perp$, the components of the dielectric tensor are 
(see, e.g., \citealt{Stix:1992})
\bea
\nonumber
D_{xx} &=& -\kpar^2 + {\omega^2\over c^2}\\
&+&\sum_s {\omega_{ps}^2\over c^2}\!\!\sum_{n=-\infty}^{+\infty}\!\!\! 
{n^2 I_n(\alpha_s)e^{-\alpha_s}\over\alpha_s}\,
\zeta_{0s}Z(\zeta_{ns}),\\
\nonumber
D_{yy} &=& D_{xx} -\kperp^2\\
&-&\sum_s {\omega_{ps}^2\over c^2}\!\!\sum_{n=-\infty}^{+\infty}\!\!\!
2\alpha_s{\partial\lt[I_n(\alpha_s)e^{-\alpha_s}\rt]\over\partial\alpha_s}\,
\zeta_{0s}Z(\zeta_{ns}),\\
\nonumber
D_{zz} &=& -\kperp^2 + {\omega^2\over c^2}\\
&-& \sum_s {\omega_{ps}^2\over c^2}\!\!\sum_{n=-\infty}^{+\infty}\!\!\!
I_n(\alpha_s)e^{-\alpha_s}
\zeta_{0s}\zeta_{ns}Z'(\zeta_{ns}),\\
\nonumber
D_{xy} &=& -D_{yx}\\ 
&=& i\sum_s {\omega_{ps}^2\over c^2}\!\!\sum_{n=-\infty}^{+\infty}\!\!\! 
n{\partial\lt[I_n(\alpha_s)e^{-\alpha_s}\rt]\over\partial\alpha_s}\,
\zeta_{0s}Z(\zeta_{ns}),\\
\nonumber
D_{xz} &=& D_{zx} = \kpar\kperp\\
&-& \sum_s {\omega_{ps}^2\over c^2}\!\!\sum_{n=-\infty}^{+\infty}\!\!\!
{n I_n(\alpha_s)e^{-\alpha_s}\over\sqrt{2\alpha_s}}\,
\zeta_{0s}Z'(\zeta_{ns}),\\
\nonumber
D_{yz} &=& -D_{zy}\\ 
&=& i\sum_s {\omega_{ps}^2\over c^2}\!\!\sum_{n=-\infty}^{+\infty}\!\!\!
\sqrt{\alpha_s\over2}{\partial\lt[I_n(\alpha_s)e^{-\alpha_s}\rt]\over\partial\alpha_s}\,
\zeta_{0s}Z'(\zeta_{ns}),
\eea
where $\alpha_s=\kperp^2\rho_s^2$,
$\zeta_{ns}=(\omega-n\Omega_s)/|\kpar|v_{{\rm th}s}$, $I_n$ are
modified Bessel functions, and $Z$ is the plasma dispersion function.

We now consider the general hot plasma dispersion ${\rm det}{\bf\hat D}=0$, or 
\bea
\nonumber
D_{xx}D_{yy}D_{zz} &+& 2D_{xy}D_{yz}D_{xz}\\
&-& D_{xz}^2D_{yy} + D_{yz}^2D_{xx} + D_{xy}^2D_{zz} = 0,
\label{disp_rln_gen}
\eea
in the asymptotic limits that we believe to be 
appropriate for the dissipation range of the solar wind: 
$\kperp\rho_i\gg1$, $\kperp\rho_e\ll1$, and $\kpar\ll\kperp$.
Then we may expand the Bessel and plasma dispersion functions 
in the above expressions for $D_{ij}$, taking $\kpar\ll \kperp$, 
$\alpha_e\ll1$, $\zeta_{0e}\ll1$, $\zeta_{ne}\gg1$ for $n\neq0$, 
$\alpha_i\gg1$, $\zeta_{ni}\gg1$. 
After straightforward algebra, we get
\bea
D_{xx} &=& -\kpar^2\lt[1 - 2\tomega^2\lt(1-C\rt)\rt],\\
D_{yy} &=& -\kperp^2\lt(1 - \beta_i{T_e\over T_i}\,L\rt),\\
D_{zz} &=& -\kperp^2\lt[1 - 2\,{T_i\over T_e}\,\tomega^2\lt(1+L\rt)\rt],\\
D_{xy} &=& i\kpar\kperp \tomega \sqrt{\beta_i},\\
D_{xz} &=& \kpar\kperp,\\
D_{yz} &=& i\kperp^2 \tomega \sqrt{\beta_i}\lt(1+L\rt),
\eea
where $\tomega=\omega/|\kpar|v_A\kperp\rho_i$.  
The (small) terms containing 
\beq
L=i\tomega\kperp\rho_i\sqrt{{\pi\over\beta_i}{T_i\over T_e}{m_e\over m_i}}
\eeq
are responsible for  electron Landau damping of the kinetic \Alfven waves, 
while the effect of the cyclotron resonance is (to lowest order) 
contained only in the matrix element $D_{xx}$ (corresponding 
to perpendicular electrostatic fluctuations): 
\bea
\nonumber
C &=& {2\over\sqrt{\pi}\kperp\rho_i}\sum_{n=1}^\infty
{\tomega^2\over\tomega^2 - n^2\tOmega_i^2}\\ 
&&- {i\over\sqrt{\beta_i}}
\sum_{n=1}^\infty {n^2\tOmega_i^2\over\tomega}
\lt(e^{-\zeta_{ni}^2} + e^{-\zeta_{-ni}^2}\rt),
\label{C_formula}
\eea
where $\tOmega_i=\Omega_i/(|\kpar v_A|\kperp\rho_i)$
and, in terms of the normalized frequencies $\tomega$ and $\tOmega_i$, 
$\zeta_{ni}=(\tomega-n\tOmega)\kperp\rho_i/\sqrt{\beta_i}$. 
The dispersion relation \exref{disp_rln_gen} can then be written 
as follows:
\bea
\nonumber
\tomega^2\lt[2 + \lt(1+{T_e\over T_i}\rt)\beta_i + (2+\beta_i)L\rt] 
- \lt(1+{T_e\over T_i}\rt)\qquad\\
\nonumber
- \lt[1-{T_e\over T_i}\lt(1+{T_e\over T_i}\rt)\beta_i\rt]L\\
= C \lt[\tomega^2\lt(2+{T_e\over T_i}\,\beta_i + 2L\rt) 
-{T_e\over T_i}\lt(1 - {T_e\over T_i}\,\beta_i L\rt)\rt].
\label{disp_rln}
\eea
If we set $C=0$ here, we recover the gyrokinetic dispersion relation 
(in the limit $\kperp\rho_i\gg1$, $\kperp\rho_e\ll1$), whose solutions 
are the weakly Landau-damped kinetic \Alfven waves (see section 2.6.2 
of \citet{Howes:2006}). The normalized frequency 
and damping rate of these waves in our notation~are 
\bea
\tomega = \tomega_{\rm KAW} &=& {1\over\sqrt{\beta_i + 2/(1+T_e/T_i)}},\\
\nonumber
\tgamma = -\tgamma_{\rm KAW} &=& 
-{\kperp\rho_i\over2}\sqrt{{\pi\over\beta_i}{T_e\over T_i}{m_e\over m_i}}\\
&& \times\lt\{1 - 2{1+ (1+T_e/T_i)\beta_i\over[2+(1+T_e/T_i)\beta_i]^2}\rt\}.
\label{gammaL}
\eea

The kinetic \Alfven waves will be affected by the ion cyclotron resonance when 
their frequency is close the cyclotron frequency, i.e., 
$\tOmega_i\sim\tomega_{\rm KAW}$, or 
\beq
\label{crossover}
{\kpar\over\kperp}(\kperp\rho_i)^2\sim 1. 
\eeq
For the highly anisotropic fluctuations in the kinetic \Alfven wave
cascade, we have $\kpar/\kperp\ll1$, so condition \exref{crossover} is
reached deep in the dissipation range, $\kperp\rho_i\gg1$---where
exactly depends on the degree of anisotropy of the fluctuations and
can be estimated using the critical balance assumption to relate
$\kpar$ and $\kperp$ (see \eqref{eq:rough}, which also includes the
dependence on $\beta_i$ and $T_i/T_e$).  Note that gyrokinetics breaks
down at this point.

Let us consider what happens in the region 
of the wavenumber space where the condition \exref{crossover} is satisfied. 
Define $\Delta_n\equiv\tomega_{\rm KAW} - n\tOmega_i$ and 
assume $\Delta_n\ll n\tOmega\sim\tomega_{\rm KAW}\sim1$.  
We now look for solutions of the dispersion relation \exref{disp_rln} 
in the form $\tomega=\tomega_{\rm KAW} + \domega$, 
where $\domega\sim\Delta_n\ll\tomega_{\rm KAW}$ 
(the validity of this assumption will be confirmed shortly). 
In the dispersion relation \exref{disp_rln}, we neglect 
the Landau damping terms (they are controlled 
by an independent small parameter, $L\propto\sqrt{m_e/m_i}$) 
and get 
\bea
\label{disp_rln_approx}
&&\tomega_{\rm KAW} \domega
= {C\over[2+(1+T_e/T_i)\beta_i]^2},\\
\label{C_approx}
&&C \simeq n\tOmega_i\lt[{1\over\sqrt{\pi}\kperp\rho_i(\Delta_n+\domega)}
- {i\over\sqrt{\beta_i}}\, e^{-\zeta_{ni}^2}\rt].
\eea
We shall see that the imaginary part of $C$ is exponentially small, 
so to lowest order, \eqsref{disp_rln_approx}{C_approx} combine 
into a quadratic equation for $\domega$. The solutions of this equation~are
\bea
\domega = -{\Delta_n\over2} \pm 
\sqrt{{\Delta_n^2\over 4} + {1\over\sqrt{\pi}\kperp\rho_i[2+(1+T_e/T_i)\beta_i]^2}}.
\label{crossover_sln}
\eea

When the kinetic \Alfven wave frequency is not too close to the 
cyclotron resonance, namely $|\Delta_n|\gg1/\sqrt{\kperp\rho_i}$, 
the ``$+$'' solution corresponds to the kinetic \Alfven wave
\beq
\tomega \simeq \tomega_{\rm KAW} + {1\over\sqrt{\pi}\kperp\rho_i\Delta_n[2+(1+T_e/T_i)\beta_i]^2}
\eeq
and the ``$-$'' solution to an approximately electrostatic wave 
called the ion Bernstein wave (see, e.g., \citealt{Stix:1992}), 
whose frequency is close to $n\Omega_i$:
\beq
\tomega \simeq n\tOmega_i - {1\over\sqrt{\pi}\kperp\rho_i\Delta_n[2+(1+T_e/T_i)\beta_i]^2}.
\eeq
(this solution is high frequency and is not captured by gyrokinetics). 

When $|\Delta_n|\sim 1/\sqrt{\kperp\rho_i}$, the more general formula 
\exref{crossover_sln} describes the conversion of the kinetic \Alfven 
wave into the ion Bernstein wave below the cyclotron resonance 
($\tomega_{\rm KAW}<\tOmega_i$) and of the ion Bernstein wave 
into the kinetic \Alfven wave above the resonance ($\tomega_{\rm KAW}>\tOmega_i$). 
This transition is illustrated in \figref{fig:app} for the $n=1$ 
resonance. The behavior at $n>1$ is similar. 

The damping rate 
can now be calculated perturbatively: restoring the Landau term $L$ 
and the imaginary part of $C$ and replacing 
$\domega\to\domega + i\tgamma$ in 
\eqsref{disp_rln_approx}{C_approx} with $\domega$ given by 
\eqref{crossover_sln}, we get (keeping only the 
lowest order terms for $\kperp\rho_i\gg1$)
\beq
\label{gamma_sln}
\tgamma = -{\Delta_n + \domega\over\Delta_n + 2\domega} 
\lt[\tgamma_{\rm KAW} + 
{1\over\sqrt{\beta_i}[2+(1+T_e/T_i)\beta_i]^2}\,e^{-\zeta_{ni}^2}\rt],
\eeq
where $\tgamma_{\rm KAW}$ is the electron Landau 
damping rate given by \eqref{gammaL}
and $\zeta_{ni}=(\Delta_n + \domega)\kperp\rho_i/\sqrt{\beta_i}$. 
Since $\domega\sim\Delta_m\sim1/\sqrt{\kperp\rho_i}$, 
we find that $\zeta_{ni}\sim\sqrt{\kperp\rho_i}$, so the 
cyclotron damping remains exponentially small and 
we have thus confirmed {\em a posteriori} that the expansion 
of $Z(\zeta_{ni})$ in large argument was justified. 
The electron Landau damping rate (the first term in \eqref{gamma_sln})
and the ion cyclotron damping rate (the second term in \eqref{gamma_sln})
are plotted in \figref{fig:app}. Note that the ion cyclotron damping 
is quite weak both in absolute terms and in comparison with 
the electron Landau damping. 

The main points that we have demonstrated by the above calculation 
are itemized at the beginning of \secref{sec:conversion}, where 
they are followed by a discussion of a somewhat less asymptotic 
case. That discussion and a much more detailed numerical study 
by \citet{LiHabbal:2001} demonstrate some of the possibilities 
for mode conversion and damping at the ion cyclotron resonances 
that are beyond the scope of a simple asymptotic calculation like 
the one presented in this Appendix. 

\end{appendix}


\end{article}

\begin{table}
\caption{Definitions}
\begin{flushleft}
\begin{tabular}{lc}
Species & $s$ \\
Speed of light & $c$ \\
Mass & $m_s$ \\
Charge & $q_s$ \\
Number Density & $n_s$ \\
Temperature\tablenotemark{a} & $T_s$ \\
Species Plasma Beta & $\beta_s = 8 \pi n_s T_s /B_0^2$ \\
Plasma Frequency & $\omega_{ps}= \sqrt{ 4 \pi n_s q_s^2 /m_s}$\\
Cyclotron Frequency & $\Omega_s = q_s B_0/(m_s c)$ \\
Thermal Velocity & $v_{{\rm th}s} =  \sqrt{2 T_s/m_s}$ \\
Alfven Velocity & $v_A= B_0\sqrt{4 \pi n_i m_i}$ \\
Mean Free Path & $\lambda_{{\rm mfp}s}$\\
Larmor Radius & $\rho_s = v_{{\rm th}s}/\Omega_s$ \\
Inertial Length & $d_s= c/\omega_{ps}$
\end{tabular}
\end{flushleft}
\tablenotetext{a}{Temperature is given in energy units, 
absorbing the Boltzmann constant.}
\label{tab:defs}
\end{table}

\begin{figure}
\resizebox{\textwidth}{!}{\includegraphics{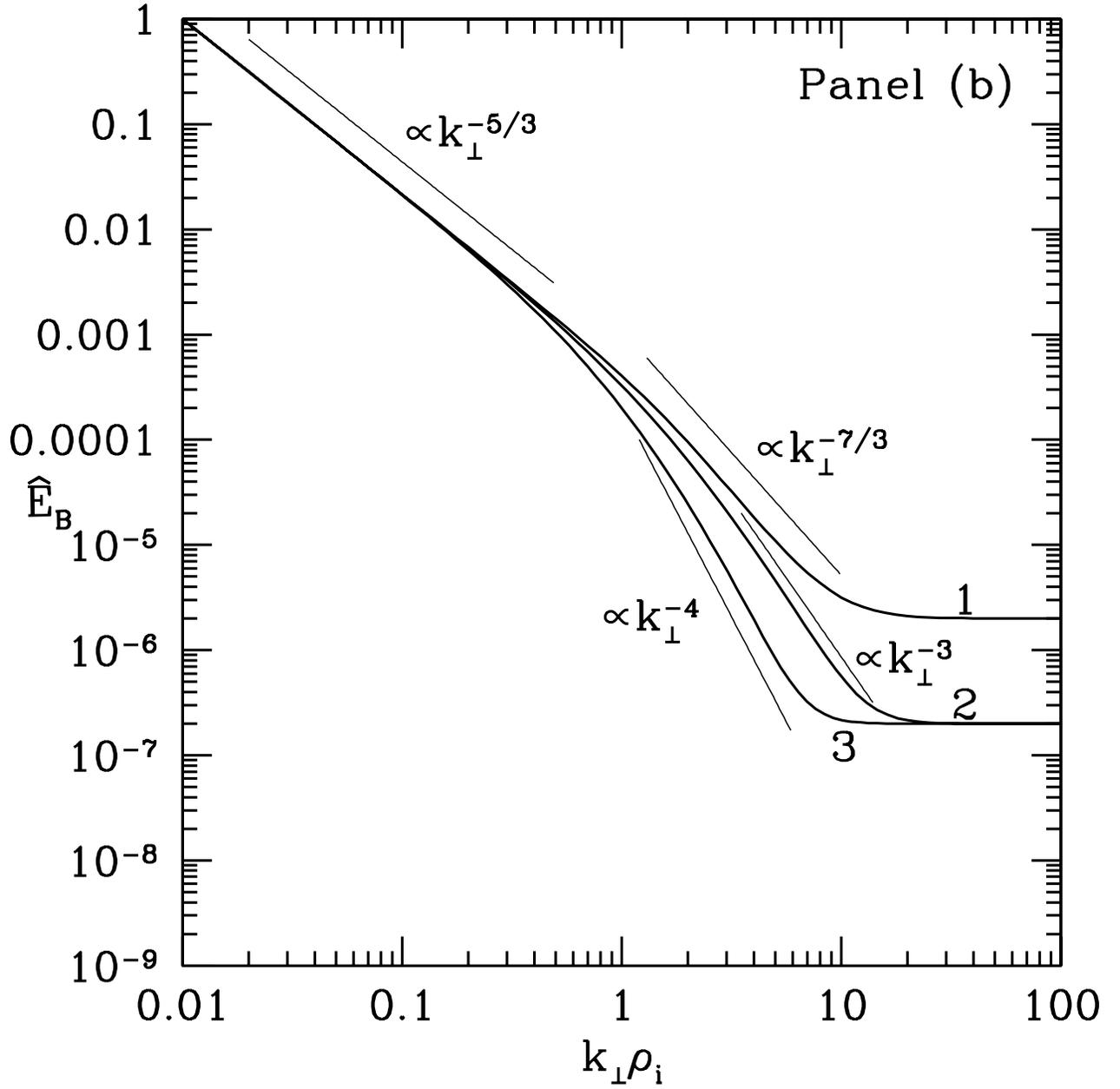}}
\caption{\label{fig:diss_range_b} One-dimensional magnetic energy
spectra for three gyrokinetic models: (1) $\beta_i=0.5$, $T_i/T_e=3$,
(2) $\beta_i=3$, $T_i/T_e=0.6$, (3) $\beta_i=0.03$, $T_i/T_e=0.175$.
Panel (b) shows the same spectra as panel (a) with an added constant
sensitivity limit, which causes resulting the spectra to resemble
power laws.  The sensitivity limit is two orders of magnitude below
the value of the spectrum at $\kperp\rho_i=1$ for spectrum 1, three
for spectra 2 and~3.}
\end{figure}

\begin{figure}
\resizebox{\textwidth}{!}{\includegraphics{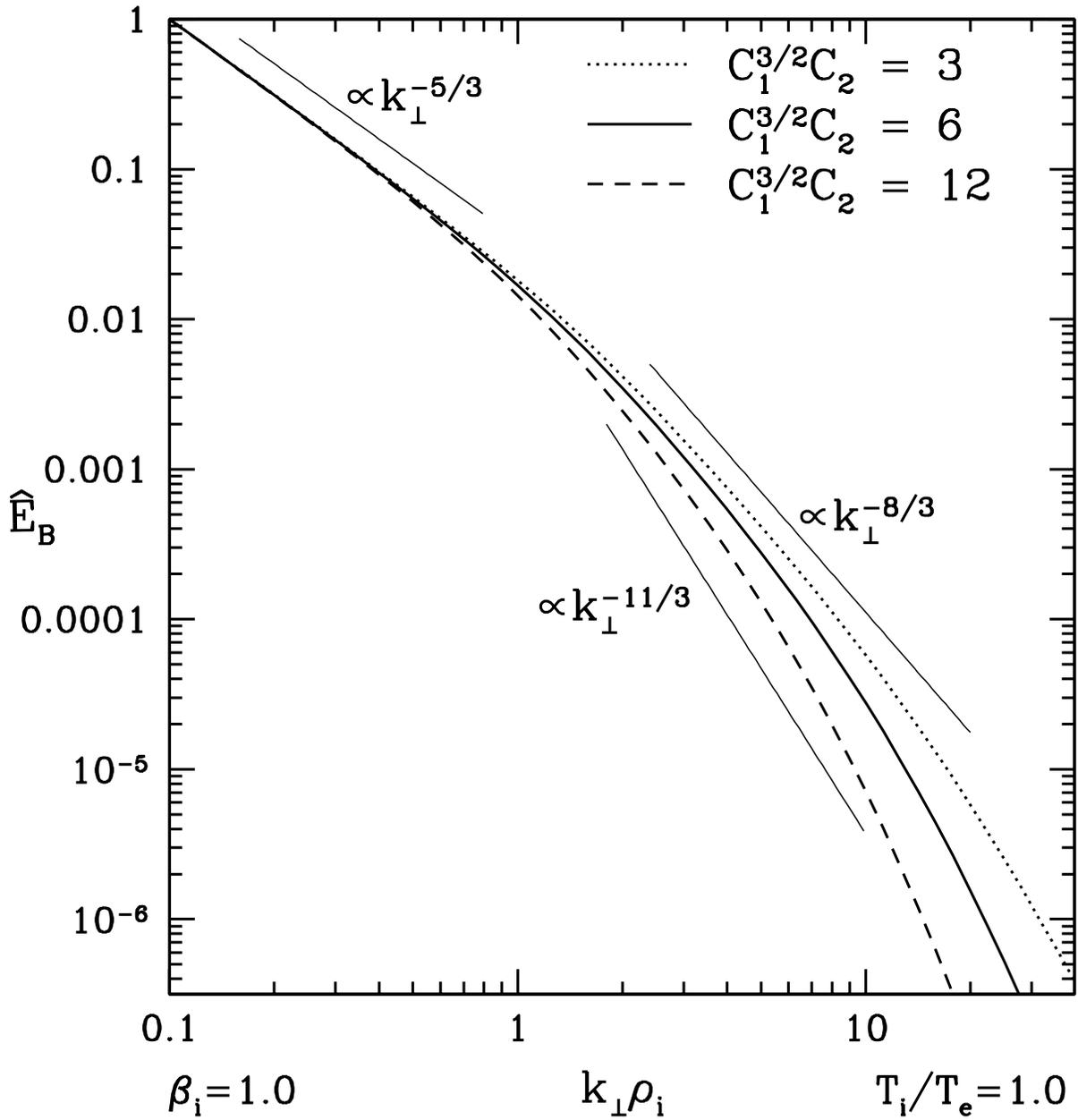}}
\caption{\label{fig:varykc}  The effect of changing the Kolmogorov
constants over a range of values $C_1^{3/2}C_2=3$, 6, and 12. 
The measured effective spectral index in the dissipation 
range varies by $+ 1/3$ or$-2/3$.}
\end{figure}

\begin{figure}
\resizebox{\textwidth}{!}{\includegraphics{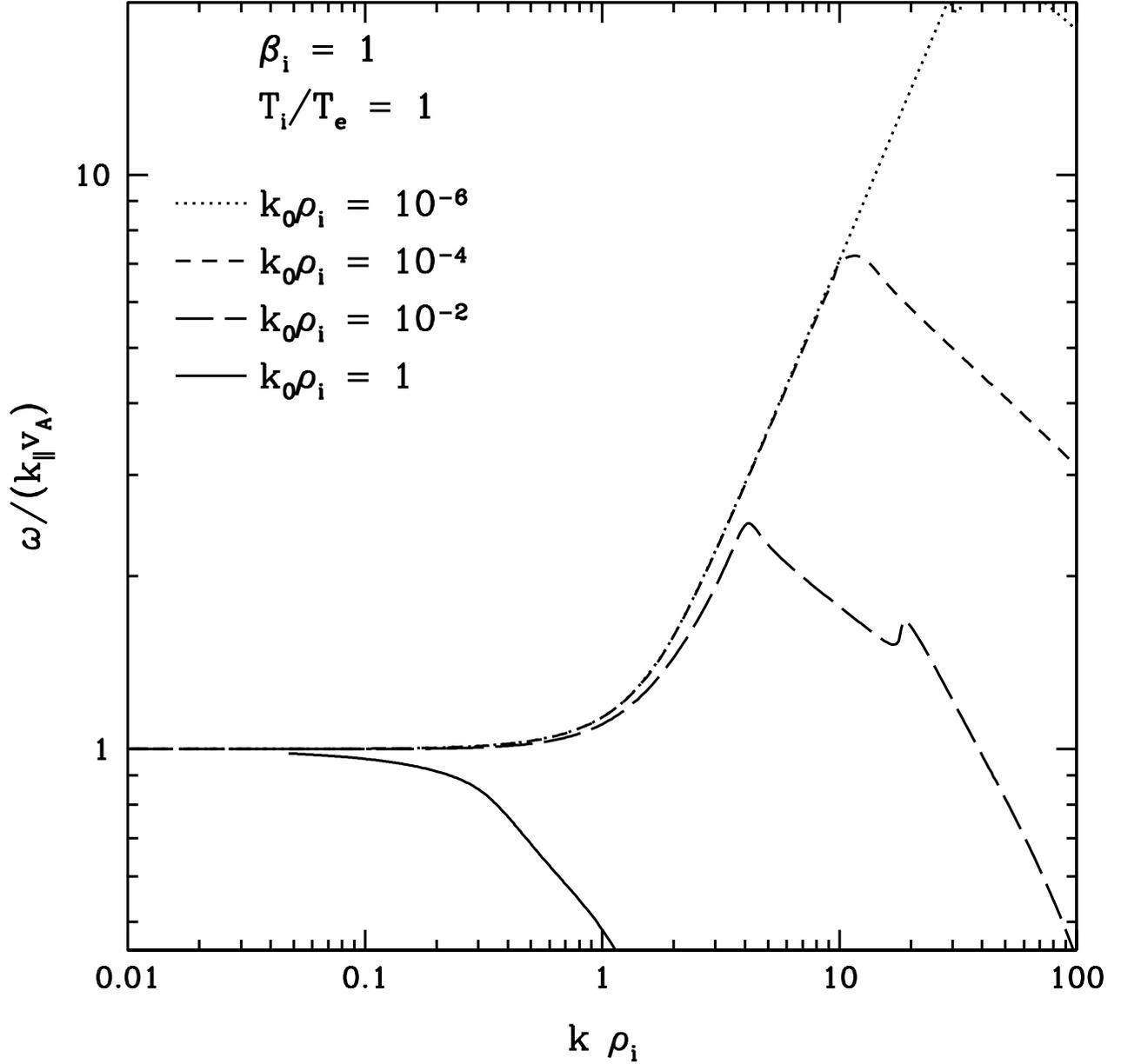}}
\caption{\label{fig:w_kpar} The normalized wave phase velocity
$\omega/(k_\parallel v_A)$ vs.~wavenumber $k \rho_i$ for models with
different assumptions about the anisotropy of the fluctuations.  The
calculations use the full linear plasma dispersion relation for a
collisionless plasma with $\beta_i = 1$ and $T_i/T_e = 1$.  We assume
a critically balanced cascade with the parallel wavenumber governed by
\eqref{eq:kpara} with no dissipation, $\epsilon = \epsilon_0$, and
driving scales $k_0\rho_i = 10^{-6}, 10^{-4}, 10^{-2}, \mbox{ and
}1$. The rise in wave phase velocity at $k \rho_i\sim 1$ for
$k_0\rho_i \ll 1$ is due to the transition to kinetic \Alfven waves;
this is consistent with \ the \emph{in situ} measurements by
\citep{Bale:2005}. The turn-down in wave phase velocity, due to the
onset of ion cyclotron damping, is not observed.}
\end{figure}

\begin{figure}
\resizebox{\textwidth}{!}{\includegraphics{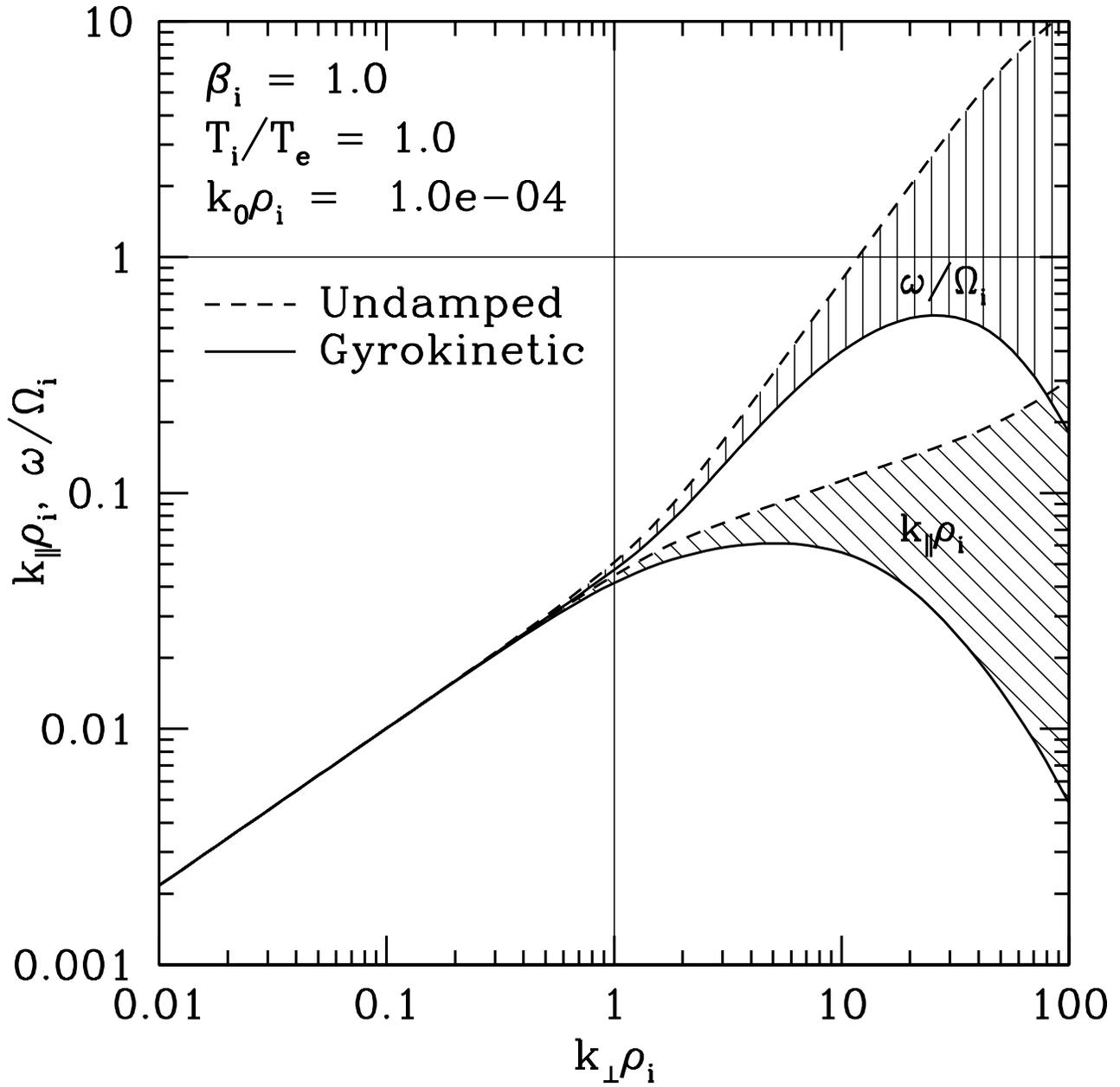}}
\caption{\label{fig:undamped} The frequency 
$\omega/\Omega_i$ [\eqref{eq:womegai2}] and parallel wavenumber
$k_\parallel \rho_i$ vs.~$k_\perp \rho_i$ [\eqref{eq:kpara}] for
undamped (dashed) and gyrokinetic (solid) steady-state solutions of
the gyrokinetic cascade model. If the damping of nonlinear turbulent
fluctuations is less than or equal to the linear gyrokinetic damping
rates, than the nonlinear frequency should lie within the vertically
shaded region and the parallel wavenumber within the diagonally shaded
region. The effective damping clearly plays a strong role in
determining whether the turbulent fluctuations reach the cyclotron
frequency, $\omega/\Omega_i\simeq 1$.}
\end{figure}

\begin{figure}
\resizebox{\textwidth}{!}{\includegraphics{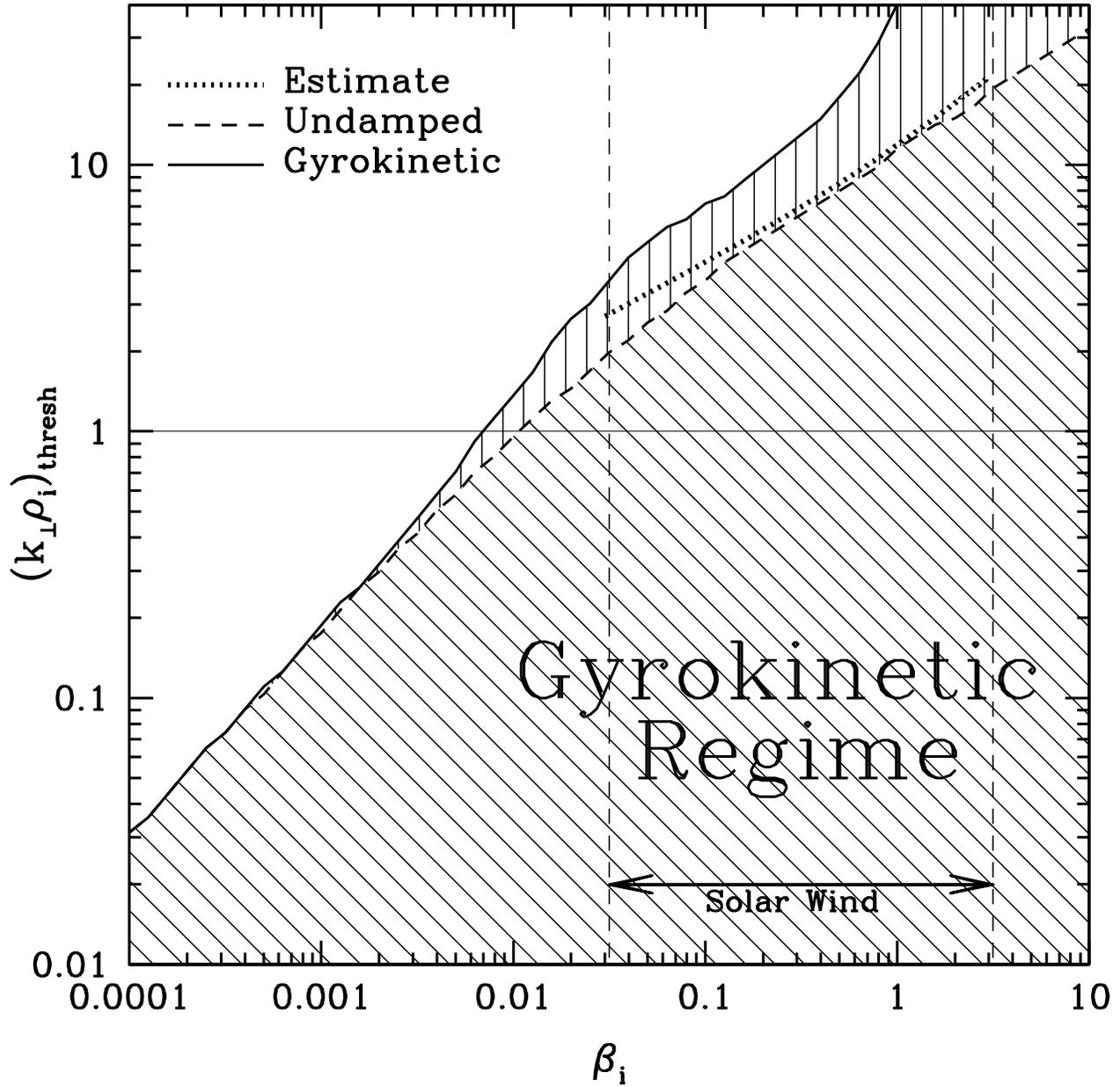}}
\caption{\label{fig:cyc_thresh_num} The threshold value of
$k_\perp \rho_i$ at which the ion cyclotron resonance has an effect on
the linear wave modes with $k_0 \rho_i=10^{-4}$ and $T_i/T_e=1$.  The
threshold is determined by comparing the solutions of the gyrokinetic
and hot plasma dispersion relations.  The solid and dashed lines
correspond to the path in the $(k_\perp,k_\parallel)$ plane calculated
from the gyrokinetic model with and without damping, respectively; the
dotted line is the analytical estimate given by \eqref{eq:rough} (only
plotted for the typical range of $\beta_i$ in the solar wind near
Earth).}
\end{figure}

\begin{figure}
\resizebox{\textwidth}{!}{\includegraphics{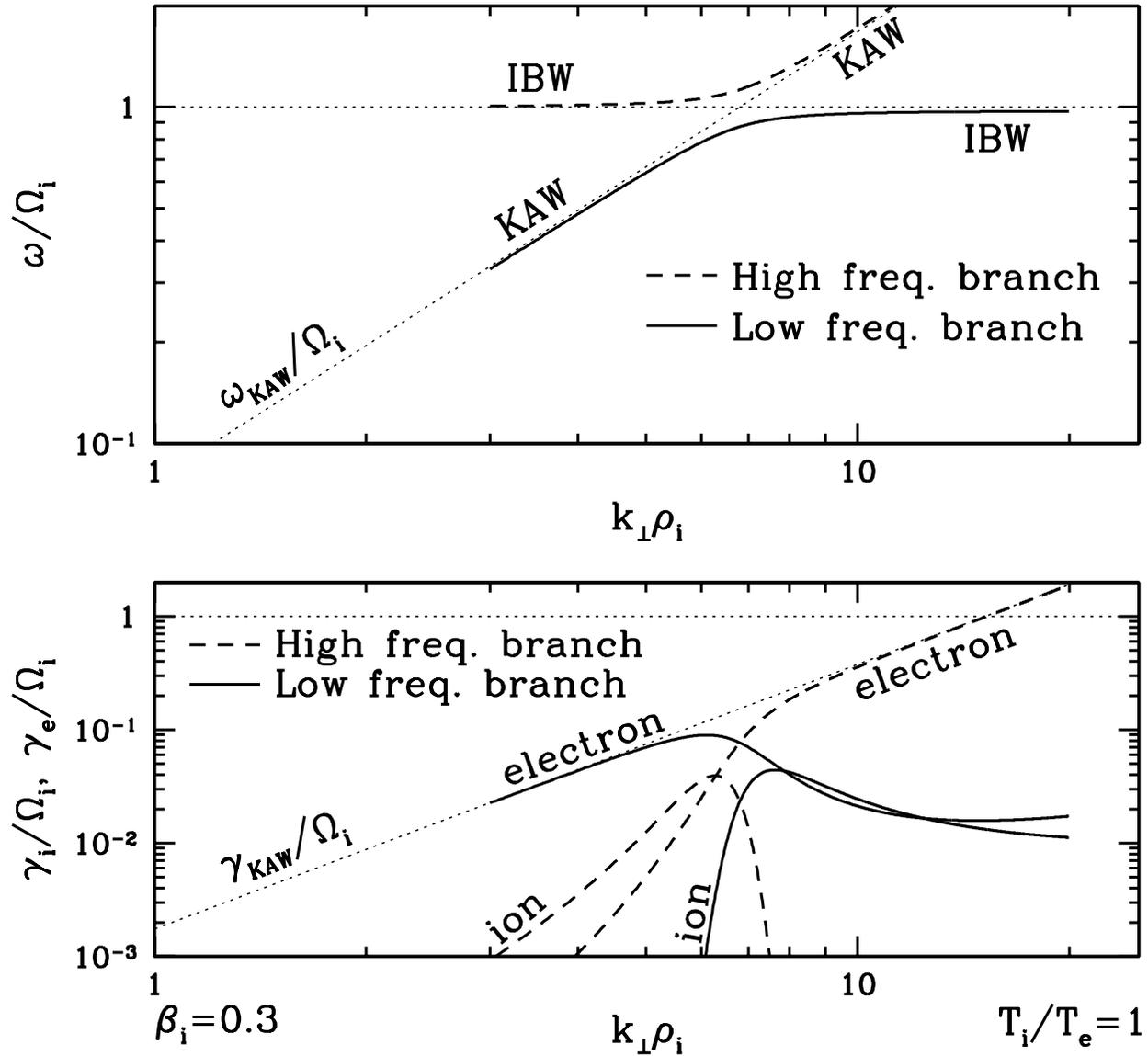}}
\caption{\label{fig:app} Approximate solution of the hot plasma 
dispersion relation illustrating the conversion between kinetic
\Alfven and ion Bernstein waves (see
Appendix~\ref{app:cycdamping}).  Plotted in the upper panel are
frequencies $\omega/\Omega_i$ [\eqref{crossover_sln}] of the low-
(solid) and high-frequency (dashed) branches. The characteristic wave
modes along each branch are labeled as kinetic \Alfven wave (KAW) and
ion Bernstein wave (IBW). Shown in the lower panel are the electron
Landau damping (electron) $\gamma_e/\Omega_i$ and ion cyclotron
damping (ion) $\gamma_i/\Omega_i$, given by the first and second terms
in \eqref{gamma_sln}, respectively. The plasma parameters are
$\beta_i=0.3$ and $T_i/T_e=1$.  The solutions are plotted along the
critically balanced cascade path with $\kpar(\kperp)$ given by
\eqref{eq:kpara}, where $k_0\rho_i=10^{-4}$ and
$\overline{\omega}=\alpha$ is given by the second 
expression in \eqref{eq:alpha}.}
\end{figure}

\begin{figure}
\resizebox{\textwidth}{!}{\includegraphics{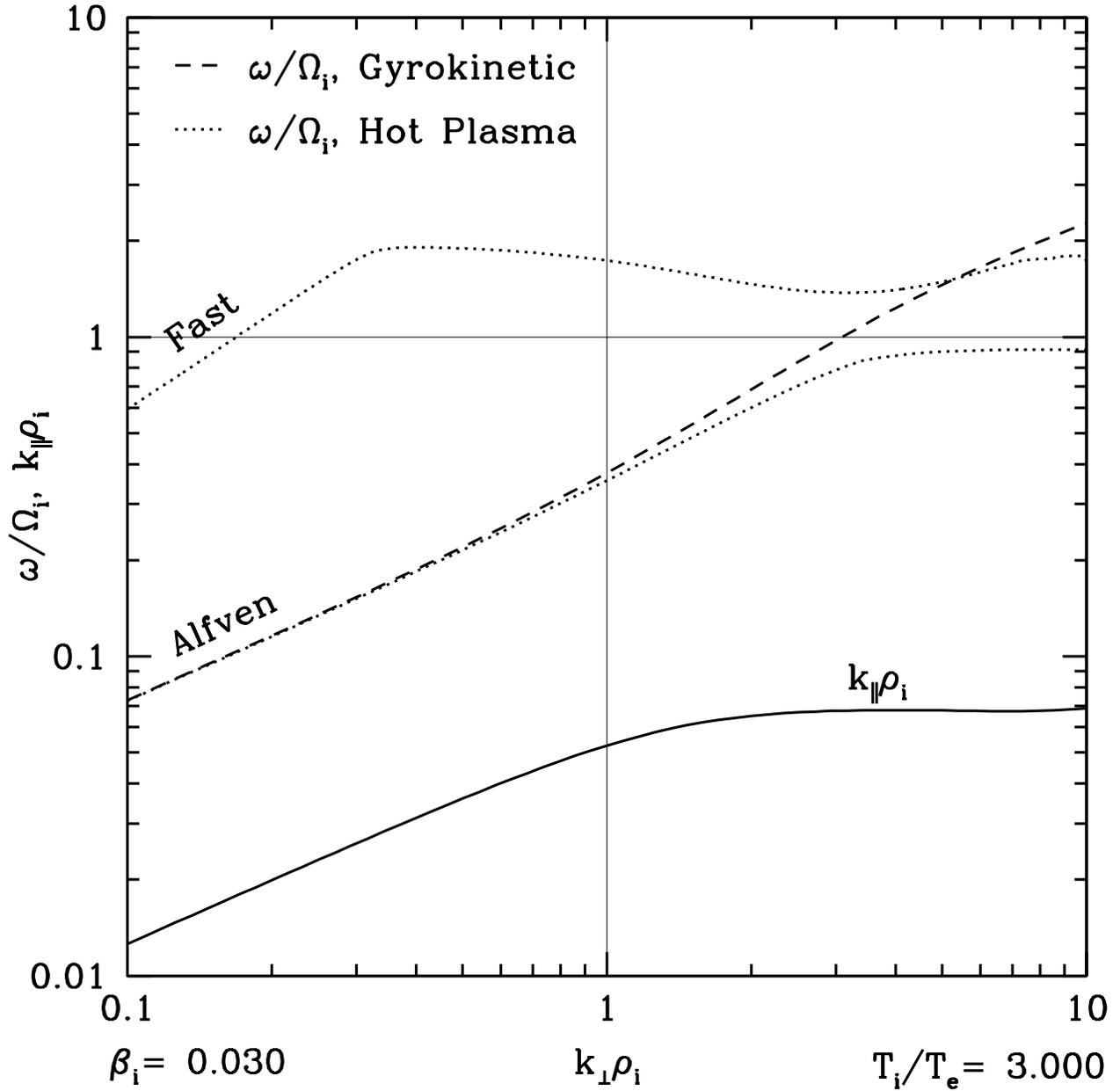}}
\caption{\label{fig:gk_hp_w_kpar} Real frequencies from the
solutions of the linear gyrokinetic dispersion relation (dashed line)
and the hot plasma dispersion relation (dotted line) for
$\beta_i=0.03$, $T_i/T_e=3$, and $k_0 \rho_i=2\times 10^{-4}$. The
solid line shows the path in the $(k_\perp,k_\parallel)$ plane implied
by critical balance. Two branches of the hot plasma solution are
shown, corresponding, in the $k_\perp \rho_i \ll1$ limit, to the MHD
\Alfven wave and the MHD fast wave.}
\end{figure}

\begin{figure}
\resizebox{\textwidth}{!}{\includegraphics{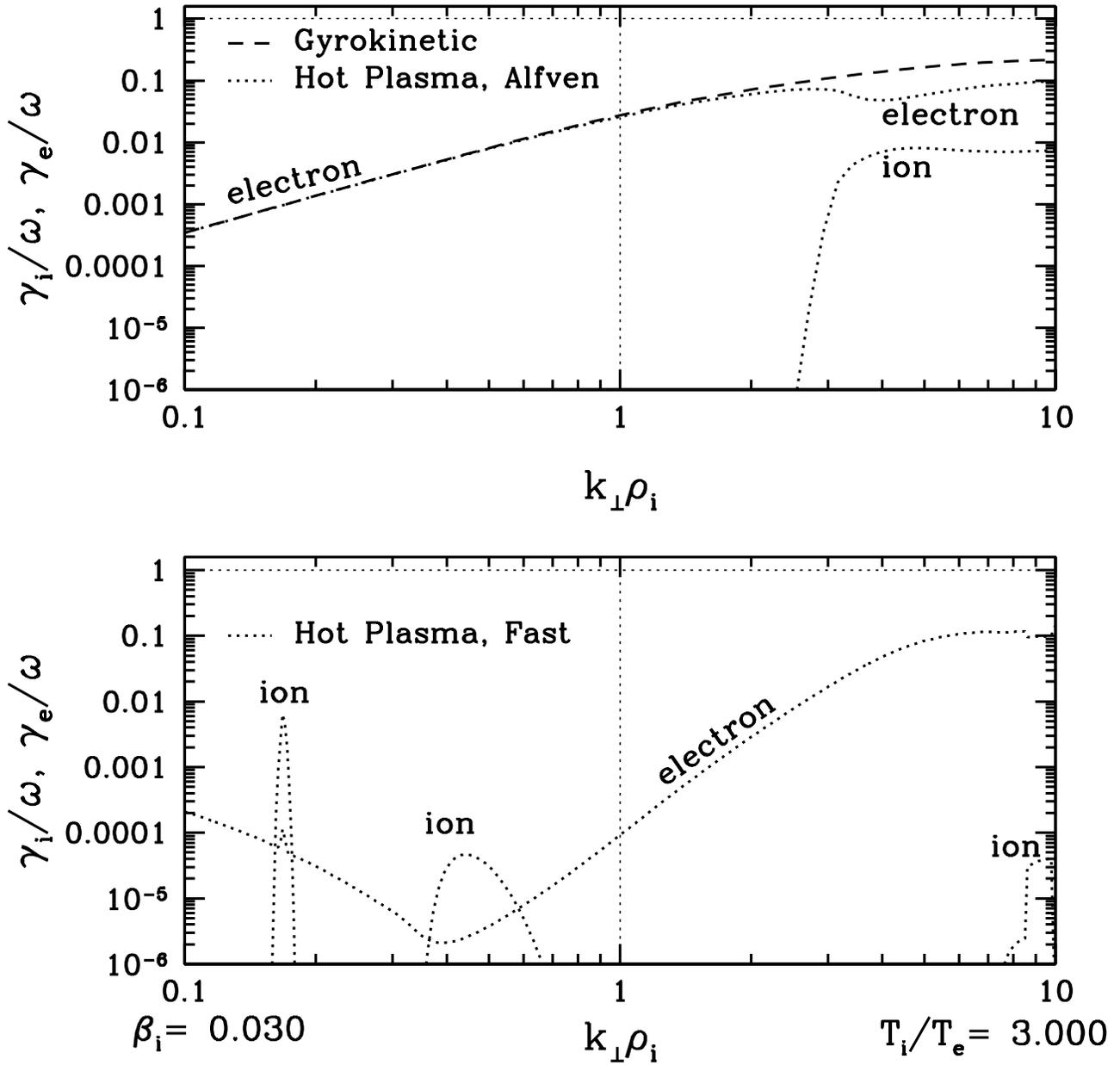}}
\caption{\label{fig:gk_hp_gwie2}
Normalized ion and electron damping rates, $\gamma_i/\omega$ and
$\gamma_e/\omega$, from the solutions of the linear gyrokinetic
dispersion relation (dashed line) and the hot plasma dispersion
relation (dotted line). In the upper panel we show the low-frequency
(Alfv\'en) branch solutions; in the lower panel, high-frequency (fast)
branch. The gyrokinetic solution shows only the electron damping---the
ion damping rates are below the plotted range. For the hot plasma
solutions, the contributions from ion and electron damping are labeled
accordingly.  These solutions are for $\beta_i=0.03$, $T_i/T_e=3$, and
$k_0 \rho_i=2\times 10^{-4}$ and the cascade path in the
$(k_\perp,k_\parallel)$ plane shown in \figref{fig:gk_hp_w_kpar}.}
\end{figure}

\begin{figure}
\resizebox{\textwidth}{!}{\includegraphics{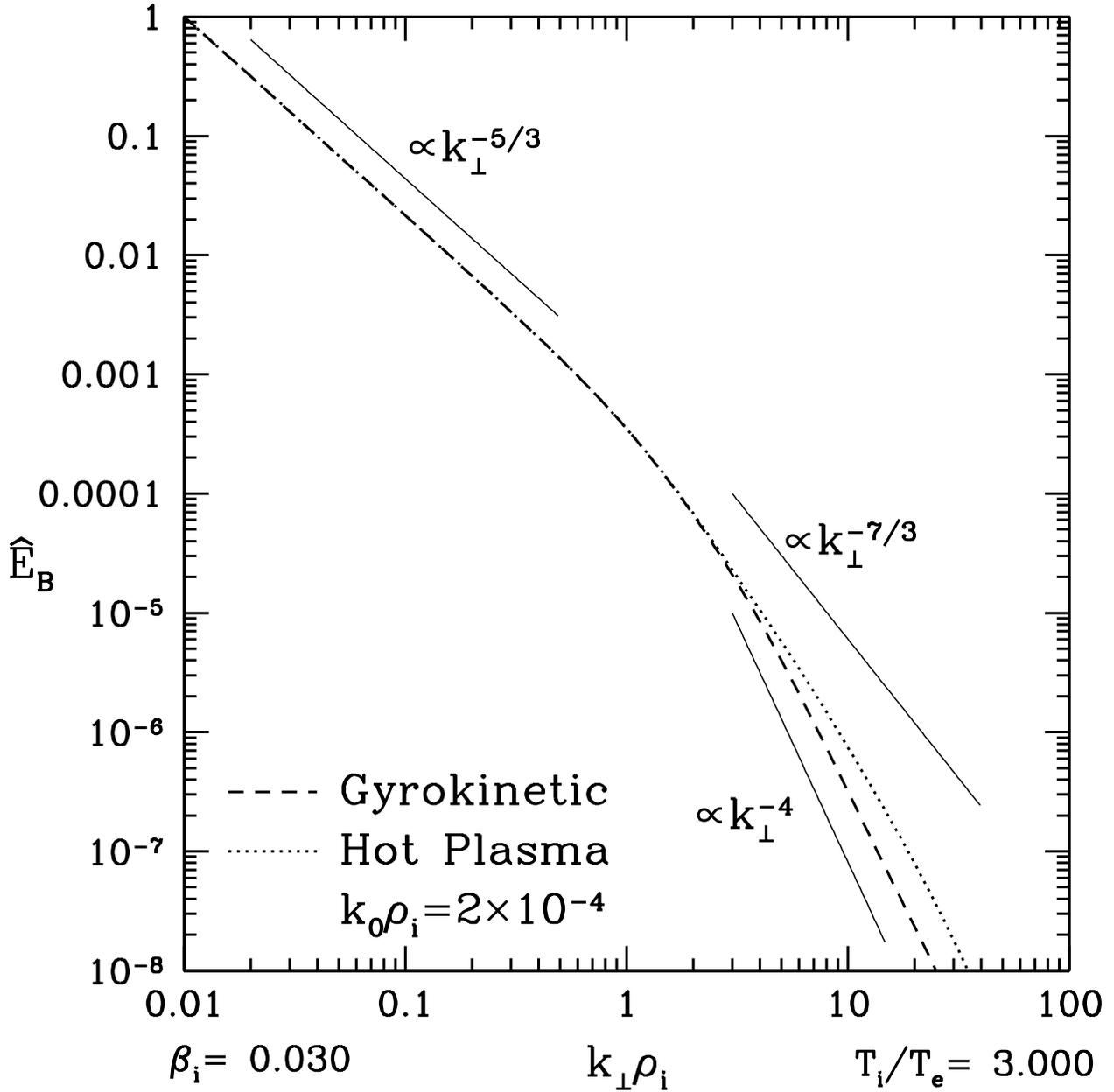}}
\caption{\label{fig:diss_gk_hp} One-dimensional magnetic energy
spectra for $\beta_i=0.03$, $T_i/T_e=3$, and $k_0 \rho_i=2\times
10^{-4}$.  Gyrokinetic and hot plasma cascade models are compared;
both models show similar spectra in the kinetic regime with 
the effective spectral
index in the dissipation range of approximately $-4$.}
\end{figure}

\begin{figure}
\resizebox{\textwidth}{!}{\includegraphics{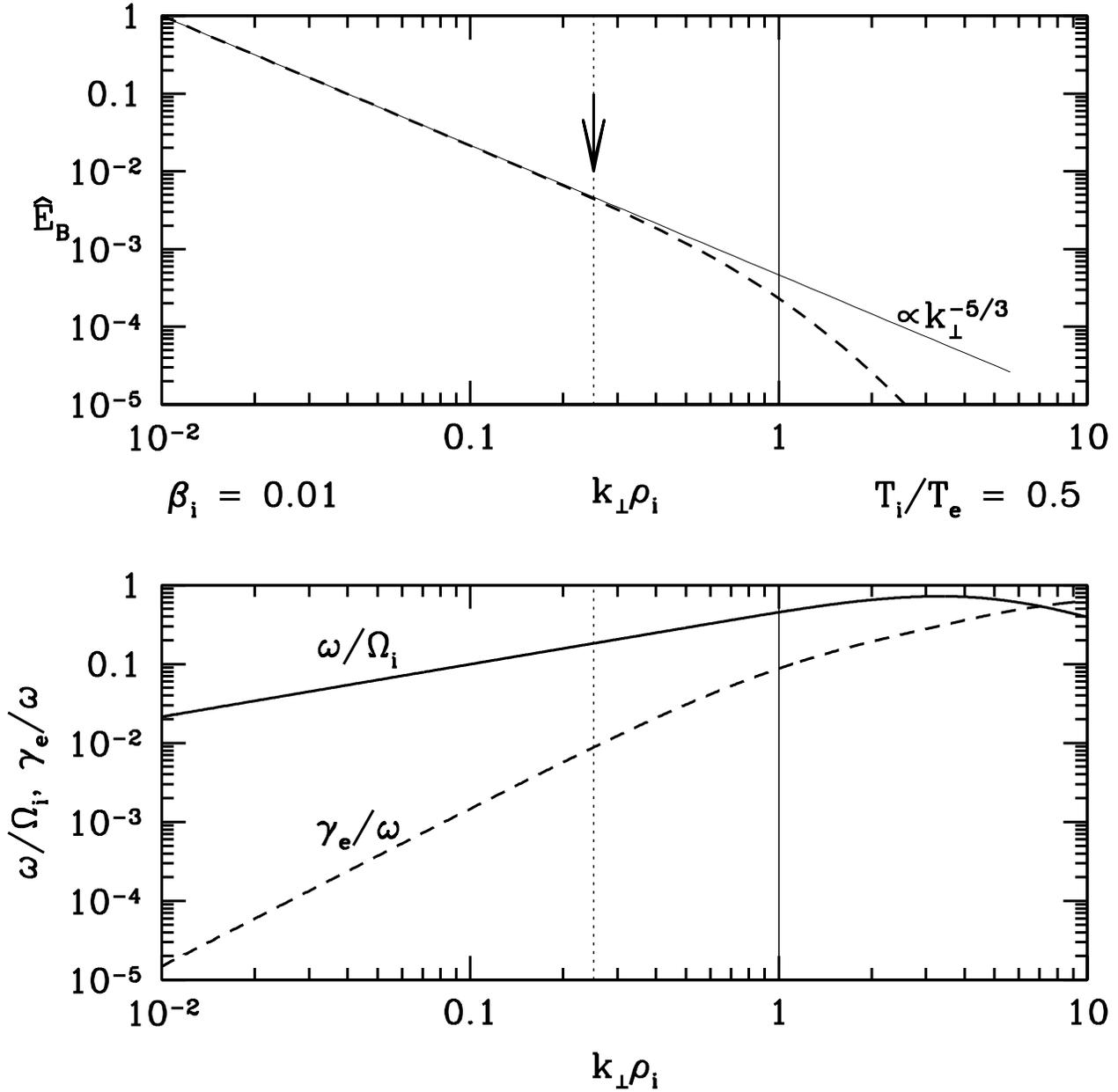}}
\caption{\label{fig:smith} The upper panel presents the normalized 
one-dimensional magnetic energy spectrum using the hot plasma cascade
model for $\beta_i=0.01$, $T_i/T_e=0.5$, and $k_0 \rho_i=10^{-4}$, the
parameters suitable for modeling the interval of solar wind studied by
\citet{Smith:2001b}. The spectrum breaks from the $k_\perp^{-5/3}$ behavior
at $k_\perp \rho_i \simeq 0.25$, as indicated by the arrow.  The lower
panel shows the frequency of the turbulent fluctuations
$\omega/\Omega_i$ and the normalized electron damping rate
$\gamma_e/\omega$ vs.~$k_\perp \rho_i$. The damping is sufficient to
prevent the cascade from reaching the ion cyclotron frequency
$\Omega_i$; therefore, the break in the spectrum at $k_\perp \rho_i
\simeq 0.25$ is due to damping via the Landau resonance with the
electrons.}  
\end{figure} 

\end{document}